\documentclass[twocolumn,prc,aps,showpacs,floatfix,amsmath,amssymb]{revtex4-2}

\usepackage{graphicx,color}
\usepackage{dcolumn}
\usepackage{bm}
\usepackage{caption} 

\usepackage{subfigure}
\usepackage{caption}
\usepackage{subcaption}
\usepackage{hyperref}
\usepackage{amsmath}
\usepackage{float}
\hypersetup{
    colorlinks,
    citecolor=blue,
    filecolor=blue,
    linkcolor=blue,
    urlcolor=blue
}

\usepackage[bottom]{footmisc}

\newcommand{\sqrtsNN}{\mbox{$\sqrt{\mathrm{s}_{_{\mathrm{NN}}}}$}}

\newcommand{\lam}{$\Lambda$}
\newcommand{\ks}{$\mathrm{K}^{0}_{\mathrm S}$}

\newcommand{\ppt}{$p_{\rm T}$}

\def \auau  {Au+Au}
\def \pbpb  {Pb+Pb}

\begin{document}

\title{Kinetic temperature and radial flow velocity estimation using identified hadrons and light (anti-)nuclei produced in relativistic heavy-ion collisions at RHIC and LHC}


\author{Junaid~Tariq}
\email{junaid.tariqhep@gmail.com}%
\affiliation{%
Department of Physics, University of Naples Federico II, University Complex of Monte Sant'Angelo Via Cinthia, 21-80126, Naples, Italy
}%

\author{M.~U.~Ashraf}
\email{usman.ashraf@cern.ch \textcolor{black}{(Corresponding author)}}%
\affiliation{%
Centre for Cosmology, Particle Physics and Phenomenology (CP3), Université Catholique de Louvain, B-1348 Louvain-la-Neuve, Belgium
}%

\author{Grigory~Nigmatkulov}
\email{gnigmat@uic.edu}%
\affiliation{%
Physics Department, University of Illinois at Chicago, Chicago, Illinois 60607, USA
}%

\date{\today}

\begin{abstract}
We report the investigation of the kinetic freeze-out properties of identified hadrons ($\pi^\pm$, $K^\pm$ and $p(\bar p)$) along with light (anti-)nuclei $d (\bar d)$, $t (\bar t)$ and ${}^{3}He$ in relativistic heavy-ion collisions at RHIC and LHC energies. A simultaneous fit is performed with the Blast-Wave (BW) model to the transverse momentum ({\ppt}) spectra of identified hadrons together with light (anti-)nuclei produced in {\auau} collisions at {\sqrtsNN} = 7.7 -- 200 GeV at the RHIC and in {\pbpb} collisions at {\sqrtsNN} = 2.76 TeV at the LHC. The energy and centrality dependence of freeze-out parameters, i.e., kinetic freeze-out temperature ($T_{kin}$) and collective flow velocity $\langle \beta \rangle$ has been studied. It is observed that light (anti-)nuclei also participate in the collective expansion of the medium created in the collision when included in a common fit with the light hadrons. We observe a marginal rise in $T_{kin}$ and a slight decrease in $\langle \beta \rangle$ when compared to the values obtained from the fit to light hadrons. A similar $\langle \beta \rangle$ and significantly larger $T_{kin}$ is observed when a fit is performed to only protons and light (anti-)nuclei. Both, $T_{kin}$ and $\langle \beta \rangle$ show a weak energy dependence at most collision energies.

\end{abstract}

\pacs{25.75.-q, 25.75.Ld, 25.75.Nq}

\maketitle

\section{Introduction}
\label{Introduction}

The investigation of the properties of nuclear matter under extreme conditions is one of the main goals of the heavy-ion program at the Relativistic Heavy Ion Collider (RHIC) and the Large Hadron Collider (LHC) experiments. The ultrarelativistic heavy-ion collisions create a condition suitable for light (anti-)nuclei production, where a significantly high energy density is achieved over a large volume.     
These conditions produce a very hot and dense matter for a very short interval of time ($\approx 10^{-22}$ seconds) containing approximately equal number of quarks and anti-quarks. This deconfined matter of quarks and gluons, the Quark-Gluon plasma (QGP), is formed during the initial state of collisions. Afterwards, the system experiences a rapid expansion with drop in its temperature and finally undergoes a phase transition towards hadron gas.

In recent years, there has been a significant emphasis on studying the production of light nuclei in heavy-ion collisions at the Bevalac~\cite{1}, SIS~\cite{2, 3}, AGS~\cite{4, 5, 6, 7, 8}, SPS~\cite{9, 10, 11, 12, 13}, RHIC~\cite{14, 15, 16, 17, 18, 19, 20, 21}, and LHC~\cite{22, 23, 24, 25, 26, 27}. The centrality and energy dependence of the production of light (anti-)nuclei has been a topic of significant interest and suggested a relevancy between the critical point and light (anti-)nuclei production~\cite{28, 29, 30, 31, 32}. However, the precise mechanisms and temporal aspects of light nuclei production in relativistic heavy-ion collisions continue to be subjects of debate due to relatively small binding energies ($\approx$ few MeV) and finite sizes of these nuclei.

It is of significant importance to study the production of light (anti-)nuclei for multiple reasons. First of all, it is not well understood how a cluster is formed in the interior of the fireball during the collision of heavy nuclei, which requires further quantitative investigation. It is probable that a substantial fraction of few-nucleon bound states observed near midrapidity are produced during the latter phases of the reaction, when the hadronic matter experiences dilution and most of the newly formed hydrogen and helium isotopes decouple from the source having no subsequent rescatterings. In this scenario, light (anti-)nuclei may be helpful to probe the dynamics of the fireball at the time of freeze-out.   

The evolution of the system produced in relativistic heavy-ion collisions is mainly described by chemical and kinetic freeze-out scenarios. The chemical freeze-out represents the stage when the inelastic collisions among hadrons stop and there is no further composition of new bound states after this stage leads to fixed yield of produced particles at this stage. At the kinetic freeze-out, the ongoing elastic interactions between produced particles results a change in the momenta of particles. The chemical and kinetic freeze-outs are assumed to happen simultaneously at the boundary between hadronic and QGP phases and the system is at chemical and kinetic equilibrium with sudden freeze-out~\cite{33}. At the other end, one might also consider another scenario, where chemical and kinetic freeze-outs take place at significantly distinct time frames. The resonances produced at the chemical freeze-out stage would decay more rapidly and the system tend to evolve with elastic collisions between hadrons. In this case, the system stays at local thermal equilibrium until the kinetic freeze-out~\cite{34}.        

The formation of (anti-)nuclei is very sensitive to conditions such as chemical freeze-out conditions, the dynamics of the emitting source and final state effects. Numerous scenarios to study the production mechanisms of (anti-)nuclei have been proposed and are typically discussed within three approaches. The first is thermal-statistical model that is very successful in describing the integrated yield of not only hadrons but also for composite nuclei~\cite{35, 36, 37, 38, 39}. The production of light (anti-)nuclei can also be studied by the coalescence model, in which protons and neutrons with similar velocities and positions on the kinetic freeze-out surface coalesce to form nuclei~\cite{40, 41, 42, 43, 44, 45, 46, 47, 48, 49}. The third approach is kinetic theory, which is based on the assumption that when light (anti-)nuclei are thermally produced at the chemical freeze-out, they might break and formed again by the final state coalescence during the evolution of the collision system~\cite{50, 51, 52, 53}. Surprisingly, it appears that both, the thermal-statistical and coalescence approaches give very consistent predictions~\cite{54}.

Additionally, the blast-wave (BW)~\cite{54a} parametrization is extensively utilized to describe the {\ppt} spectra of light (anti-)nuclei~\cite{19, 21}. The conceptual basis for the BW model is derived from its similarity to the freeze-out configuration within the hydrodynamic model~\cite{34}. The BW parametrization provides a remarkably good fit to the {\ppt} spectra of light (anti-)nuclei produced in {\auau} collisions at {\sqrtsNN} = 7.7 -- 200~GeV at RHIC~\cite{19} and {\pbpb} collisions at {\sqrtsNN} = 2.76~TeV at LHC~\cite{21}. The only difference between the two different types of collision systems is the exponent $n$ in the velocity profile ($\beta = \beta_s (\frac{r}{R})^n$), which was set to 1 at RHIC, while it was treated as a free parameter at LHC.

Many studies have been performed to investigate the energy and centrality dependence of kinetic freeze-out temperature ($T_{kin}$) and the radial flow velocity ($\beta_T$) parameters by simultaneously fitting the {\ppt} spectra of individual light hadrons (($\pi^\pm$, $K^\pm$ and $p(\bar p)$))~\cite{55, 56} produced in heavy ion collisions as well as combined with strange hadrons ({\ks} and {\lam})~\cite{ 57, 58}, and with $d, {}^{3}He$ with the BW model~\cite{21}. During the expansion phase of the strongly interacting system, the pressure gradient gives a rise to the collective flow. This results in a characteristic dependence of the shape of the {\ppt} distribution on the particle mass and can be described with a common $T_{kin}$ and $\beta_T$~\cite{54a}. By a thorough analysis on the simultaneous fit to the {\ppt} spectra of light hadrons ($\pi^\pm$, $K^\pm$ and $p(\bar p)$) along with light (anti-)nuclei ($d (\bar d)$ and $t (\bar t)$), we attempt to investigate the freeze-out parameters ($T_{kin}$ and $\beta_T$) extracted from the experimental data available for RHIC and LHC energies. In this work, we performed a systematic study of energy and centrality dependence of $T_{kin}$ and $\beta_T$ extracted from fitting the {\ppt} spectra of light hadrons along with light (anti-)nuclei with the BW model from {\auau} collisions at {\sqrtsNN} = 7.7 -- 200~GeV and {\pbpb} collisions at 2.76~TeV.

This paper is organized as follows. In Section~\ref{sec2}, we describe the analysis method. Results and Discussion are presented in Section~\ref{sec3}. The conclusion is summarized in Section~\ref{Conclusion}.

 \section{Analysis Method}
 \label{sec2}

Blast-Wave (BW)~\cite{54a} is a phenomenological model commonly used to fit the hadron as well as light (anti-)nuclei spectra by experimental collaborations~\cite{19, 21}. The model is based on the assumption that particles are emitted thermally for an expanding source with global variables: temperature ($T$) and velocity profile ($\beta$~\cite{54a, 59}). The functional form can be written as:
\begin{equation}\label{eq1}
   \frac{d^2N}{p_Tdp_T} \propto 
    \int_{0}^R r dr m_{T} I_{0}(\frac{p_T\sinh\rho(r)}{T_{kin}})\times K_1 (\frac{m_T\cosh\rho(r)}{T_{kin}}),
\end{equation}
where $m_T = \sqrt{p_{T}^2 + m^2}$ is the transverse mass of the hadron species. The $I_{0}$ and $K_1$ are the modified Bessel functions. The temperature of the localized thermal sources, from which particles radiate, is denoted by $T_{kin}$. While assuming the longitudinal expansion of the system to be boost invariant, the velocity profile of the thermal source can be parameterized as:

\begin{equation}
    \rho = tanh^{-1} \beta (r)= tanh^{-1} (\beta_S (\frac{r}{R})^n ).
\end{equation}

The radial distance from the center of the fireball in the transverse plane is represented by $r$, the radius of the fireball is $R$, the transverse expansion velocity is $\beta (r)$ at radius $0 \leq r \leq R$. The $\beta_S$ is the transverse expansion velocity at the surface and exponent of the velocity profile is $n$. The average radial flow velocity can be expressed as $\langle \beta \rangle = \beta_S~. ~2 / (2+n)$~\cite{60}. In order to study the light (anti-)nuclei distribution in different colliding systems, $n$ was treated differently. In {\auau} collisions at {\sqrtsNN} = 7.7 -- 200~GeV at RHIC~\cite{19}, $n$ was fixed to 1, while treated as a free parameter in {\pbpb} collisions at {\sqrtsNN} = 2.76 TeV at LHC~\cite{21} for better fit. This different treatment of $n$ to study the light (anti-)nuclei {\ppt} distribution at RHIC and LHC reflects the different dynamical evolution of the fireball formed in heavy ion collisions. In the current work, to check the universality of the BW parametrization we use $n=1$ for {\auau} and {\pbpb} collisions. Therefore, the free parameters are only $T$, $\beta$ and the normalization factor. Generally, the important parameters to determine the shape of each spectrum are the $T$, $n$, $\beta$ and the mass of the particle specie, but with the common $T$ and $\beta$, the mass of particle specie is an important parameter which determine the shape of the spectrum in BW model~\cite{61}. 

 \section{Results and Discussion}
\label{sec3}
The results of our analysis are presented in this section. 
\subsection{Transverse Momentum ({\ppt}) Spectra}

It is crucial to observe that blast-wave (BW) fits represent a simplified method that imitates the hydrodynamics involved in radial expansion and comes with specific constraints. For instance, it is acknowledged that the temperature is notably influenced by the chosen fit range and the particle species. Particularly, the quality of the BW fits using the FastReso package~\cite{73} is relatively good using a single temperature of $\approx$ 150~MeV for chemical and kinetic freeze-out and it has been reported in Ref.~\cite{74} that the temperature from these fits exhibits a weak dependence on centrality and collision system. This is possible in the FastReco approach through its explicit consideration of the feed-down effect from resonance decays. Furthermore, alternative methods utilize an extended BW model approach with additional parameters employed to LHC data for a better description~\cite{66a}. The conventional (Boltzmann-Gibbs) blast-wave parameterization provides a straightforward and robust method for comparing the spectra of hadrons and light (anti-)nuclei, which is the objective of the current study.

The measured {\ppt} spectra of light hadrons ($\pi^\pm$, $K^\pm$ and $p(\bar p)$) and light (anti-)nuclei ($d (\bar d)$ and $t (\bar t)$) in {\auau} collisions at {\sqrtsNN} = 7.7 -- 200~GeV from RHIC~\cite{19, 55, 62} and {\pbpb} collisions at {\sqrtsNN} = 2.76~TeV from LHC~\cite{21, 56} at available centrality classes are analyzed for this purpose. No data available for $\bar t$ in {\auau} collisions at {\sqrtsNN} = 7.7 -- 200~GeV and {\pbpb} collisions at {\sqrtsNN} = 2.76~TeV. Therefore, we use ${}^{3}He$ along with $\pi^\pm$, $K^\pm$ and $p(\bar p)$ in {\pbpb} collisions at {\sqrtsNN} = 2.76~TeV for our analysis. Details of the data sets of light hadrons and light (anti-)nuclei under investigation are summarized in Table~\ref{table1}. The simultaneous BW fits to the {\ppt} spectra of particle species under investigation are performed by minimizing the value of $\chi^2/N_{DOF}$, where $N_{DOF}$ is the number of degrees of freedom and is defined as the number of data points minus the number of fitting parameters. The minimization procedure in the current analysis is done by using the MINUIT~\cite{64} package available within the ROOT framework~\cite{65,79}.

For light hadrons, we repeated the analysis presented in Ref.~\cite{55} for a simultaneous fit to the {\ppt} spectra with BW model. The obtained parameters from the BW fit are consistent with the previously reported results~\cite{55}. The low {\ppt} part of the pion spectra is significantly influenced by resonance decays therefore, {\ppt} $< 0.5$ GeV/$c$ is excluded from the fit. Conversely, at higher {\ppt}, eventual hard contributions may become significant, which are not anticipated to be effectively described by the BW model~\cite{66a}. Therefore, the sensitivity of the BW model parameters especially $T_{kin}$ is contingent upon the choice of {\ppt} ranges used for fitting~\cite{56}. The fit range for light hadrons have been limited according to Ref.~\cite{55} (0.5 -- 1.3~GeV/$c$, 0.25 -- 1.4~GeV/$c$, 0.4 -- 1.3~GeV/$c$ for $\pi^\pm, K^\pm$ and $p(\bar  p)$ respectively). However, for light (anti-)nuclei the spectrum is fitted up to the maximum available {\ppt} range. The fit ranges can significantly affect the extracted $T_{kin}$, especially at lower {\ppt}. Further details on fit ranges are reported in Refs.~\cite{55a, 55b}.

\begin{figure}[!ht]
\includegraphics[width=0.45\textwidth]{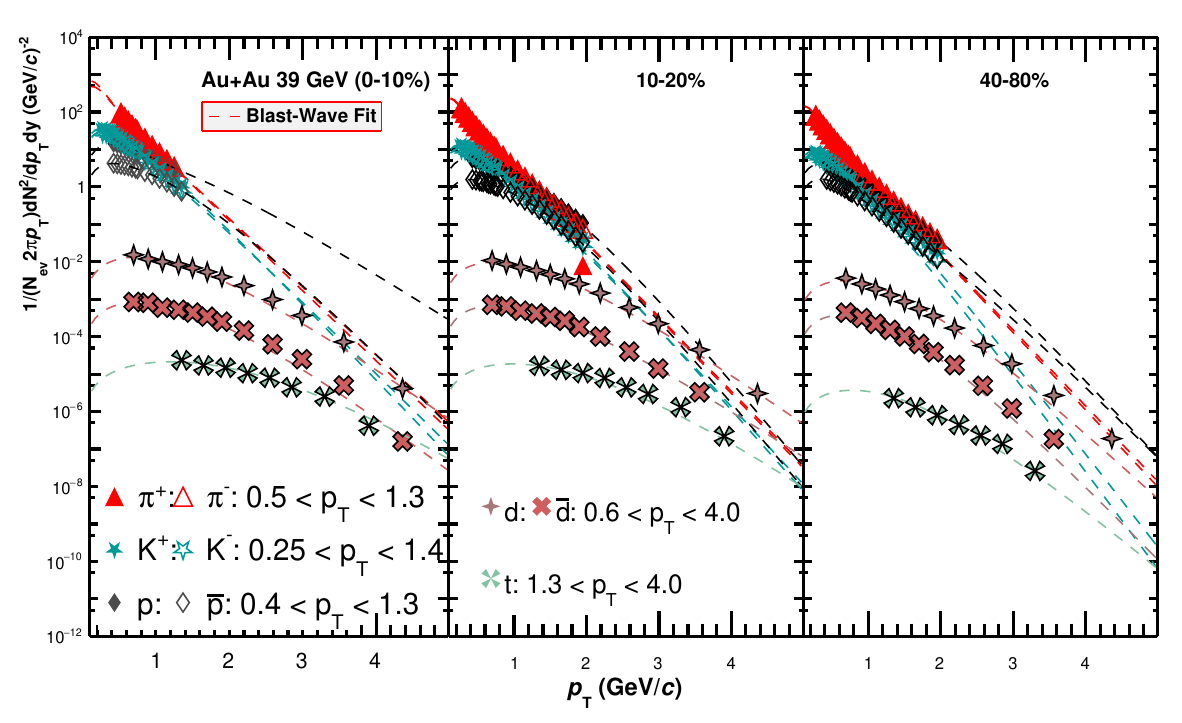}
\caption{(Color online) Blast-Wave model simultaneous fit to the {\ppt} spectra of $\pi^\pm$, $K^\pm$, $p(\bar p), d (\bar d)$ and $t (\bar t)$ in {\auau} collisions at {\sqrtsNN} = 39~GeV in 0--10\%, 10--20\% and 40--80\% centrality classes. For data points, statistical and systematic errors are added in quadrature.}
\label{fig1}
\end{figure}

\begin{figure}[!ht]
\centering
\includegraphics[width=0.45\textwidth]{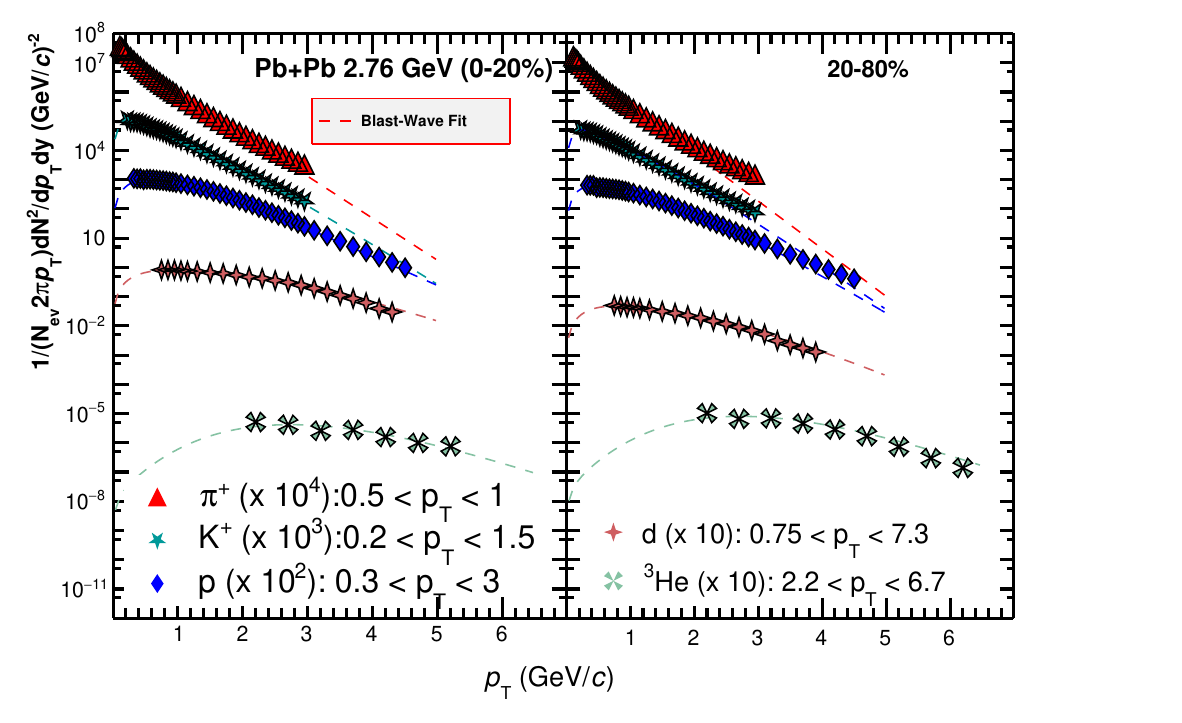}

\caption{(Color online) Blast-Wave model simultaneous fit to the {\ppt} spectra of $\pi^\pm$, $K^\pm$, $p(\bar p), d (\bar d)$ and $t (\bar t)$ in {\auau} collisions at {\sqrtsNN} = 2.76~TeV in 0--20\% and 20--80\% centrality classes. For data points, statistical and systematic errors are added in quadrature.}
\label{fig2}
\end{figure}

Figures~\ref{fig1} and ~\ref{fig2} show the simultaneous BW fit using Eq.~\ref{eq1} to the {\ppt} spectra of light hadrons and light (anti-)nuclei in {\auau} collisions at {\sqrtsNN} = 39~GeV in 0--10\%, 10--20\% and 40--80\% centrality classes and {\pbpb} collisions at {\sqrtsNN} = 2.76~TeV in 0--20\% and 20--80\% centrality intervals respectively as an example. The spectra of light (anti-)nuclei are evidently well described by the common BW fit within uncertainties indicating that relatively heavy particles, like $d (\bar d), t (\bar t)$ and ${}^{3}He$, also participate in a collective flow field. The extracted fit parameters and $\chi^2/N_{DOF}$ are tabulated in Section~\ref{Appen}.

\begin{table*}[hbt!] 
\scriptsize{
\caption{References for data}
\vspace{-.50cm}
\begin{center}
\begin{tabular}{p{2cm}p{4cm}p{3cm}p{2cm}p{2cm}}\\ \hline\hline
     System & $\sqrt{S_{NN}}$ (GeV)  &   Particle &    Collaboration & Reference  \\\hline
    {\auau}  & 7.7, 11.5, 19.6, 27, 39 & $\pi^{\pm}$, $K^{\pm}$, $p$, $\bar{p}$   & STAR & \cite{55} \\
    \hline
    {\auau}  & 14.5 & $\pi^{\pm}$, $K^{\pm}$, $p$, $\bar{p}$   & STAR & \cite{67} \\
    \hline
    {\auau}  & 62.4 & $\pi^{\pm}$,$p$, $\bar{p}$   & STAR & \cite{68} \\
    \hline
    {\auau}  & 62.4 & $K^{\pm}$   & STAR & \cite{69} \\
    \hline
    {\auau}  & 200 & $\pi^{\pm}$,$p$, $\bar{p}$   & STAR & \cite{68} \\
    \hline
    {\auau}  & 200 & $K^{\pm}$   & PHENIX & \cite{70} \\
    \hline
    {\auau}  & 7.7, 11.5, 14.5, 19.6, 27, 39, 200 & $d$, $\bar{d}$   & STAR & \cite{19} \\
    \hline
    {\auau}  & 7.7, 11.5, 14.5, 19.6, 27, 39, 200 & $t$   & STAR & \cite{62} \\
    \hline
    {\pbpb}  &2760 & $\pi^{\pm}$,$K^{\pm}p$, $\bar{p}$    & ALICE & \cite{56} \\
    \hline
    {\pbpb}  &2760 & $d$, $^3He$    & ALICE & \cite{21} \\
    \hline
    {\pbpb}  &5020 & $\pi^{\pm}$,$K^{\pm}p$, $\bar{p}$  & ALICE & \cite{72}\\
    \hline
    {\pbpb}  &5020 &  $d$,$\bar{d}$, $^3He$    & ALICE & \cite{71} \\
    \hline
    \hline

\end{tabular}
\label{table1}
\end{center}} 
\end{table*}

\subsection{Kinetic Freeze-out Parameters}

The extracted freeze-out parameters from the BW fit have been presented in this section. It is important to point out that the whole measured {\ppt} range for all particle species is very well described by the BW function (individual fit only), however, no physics meaning is attached to the parameters. A combined BW fits to different particle species can be useful to provide valuable insights into the freeze-out conditions and allow to compare quantitatively between different systems colliding at different {\sqrtsNN} in terms of hydrodynamic interpretation. In principle different particle species can decouple at different times leading to different values of $\langle \beta \rangle$ and $T_{kin}$ from the hadronic medium due to different hadronic cross sections. The interpretation of this statement becomes more arguable by forcing all the particle species to decouple assuming same set of parameters. 


\begin{figure*}
    \centering
    \begin{subfigure}
        \centering
        \includegraphics[width=0.335\linewidth, height=0.3\textwidth]{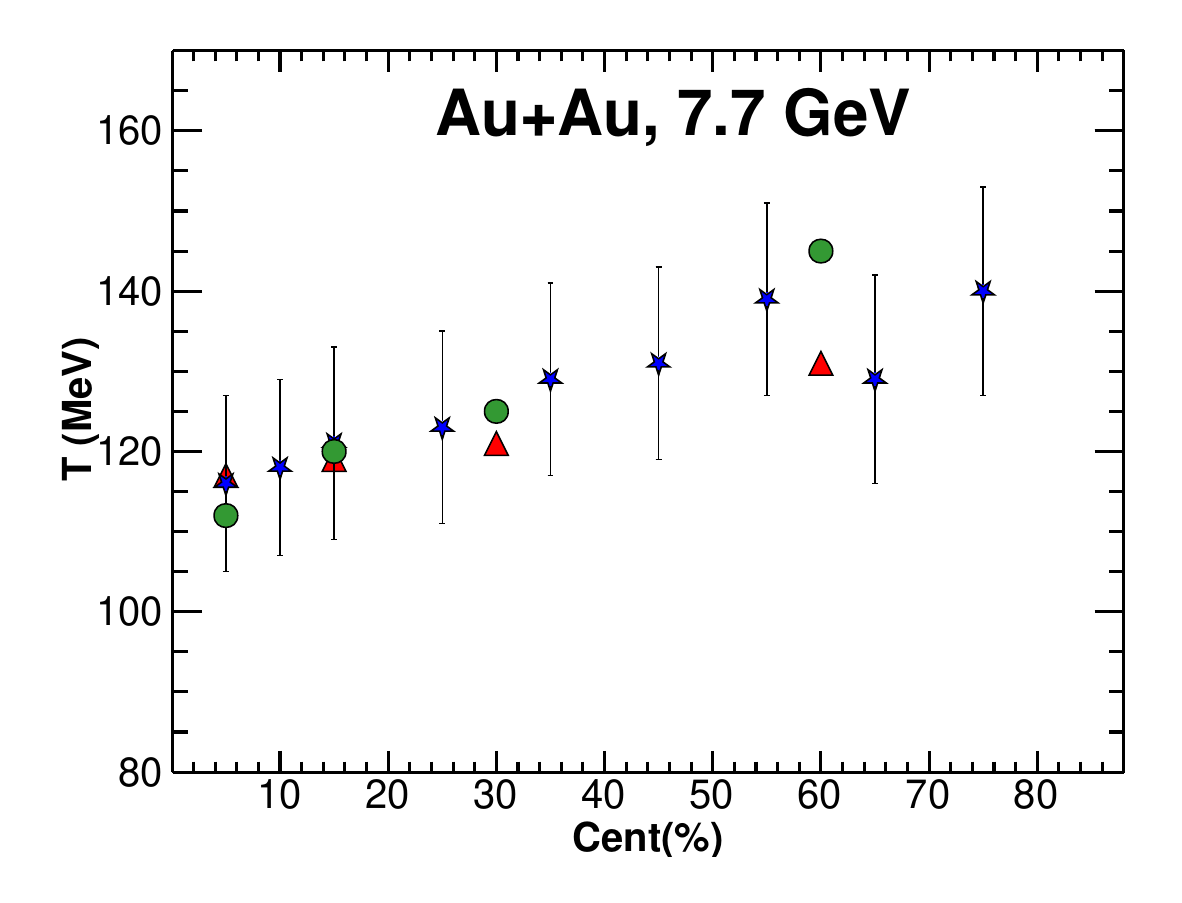} 
        \hspace{-0.5cm}
        \vspace{-0.5cm}
        \includegraphics[width=0.335\linewidth, height=0.3\textwidth]{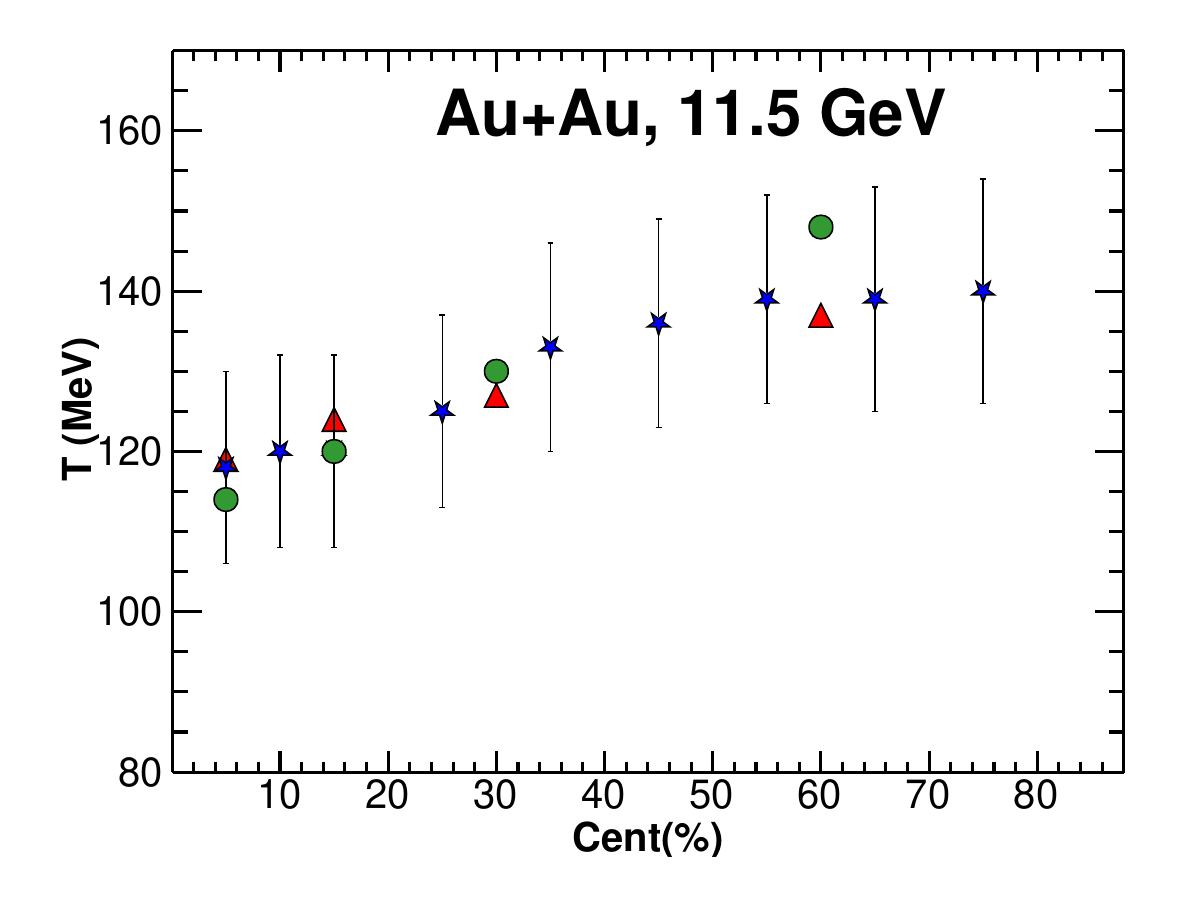}
        \hspace{-0.5cm}
        \includegraphics[width=0.335\linewidth, height=0.3\textwidth]{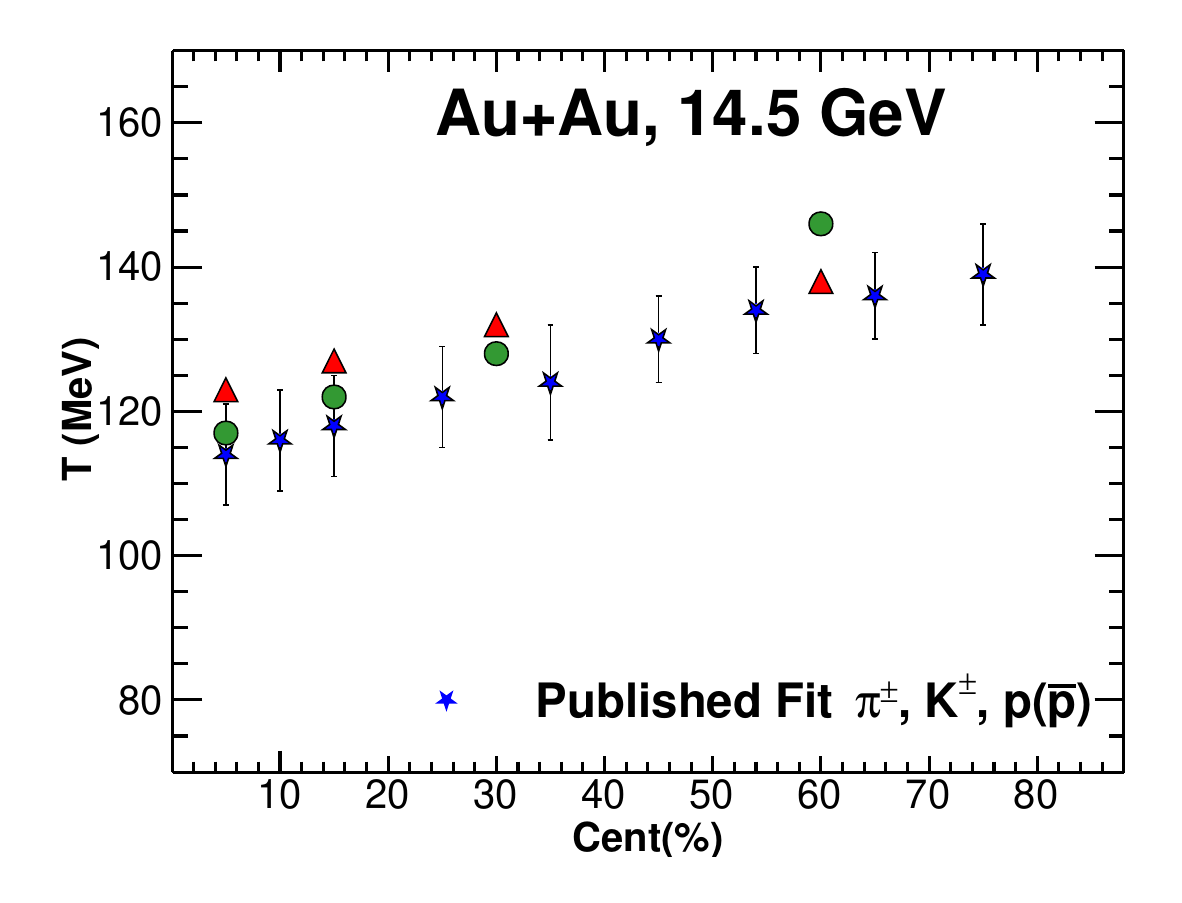}
        \hspace{-0.5cm}
        \vspace{-0.3cm}
        \includegraphics[width=0.335\linewidth, height=0.3\textwidth]{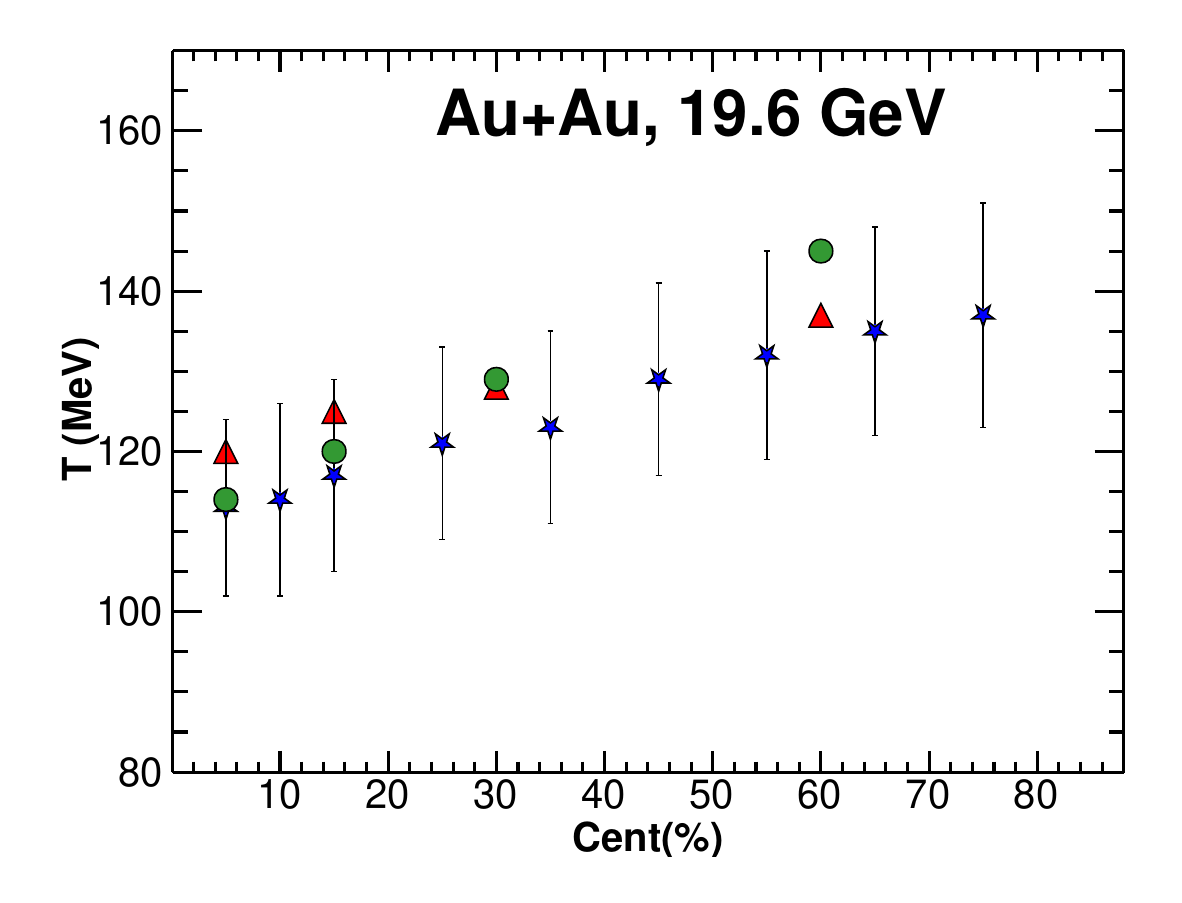}
        \hspace{-0.5cm}
        \includegraphics[width=0.335\linewidth, height=0.3\textwidth]{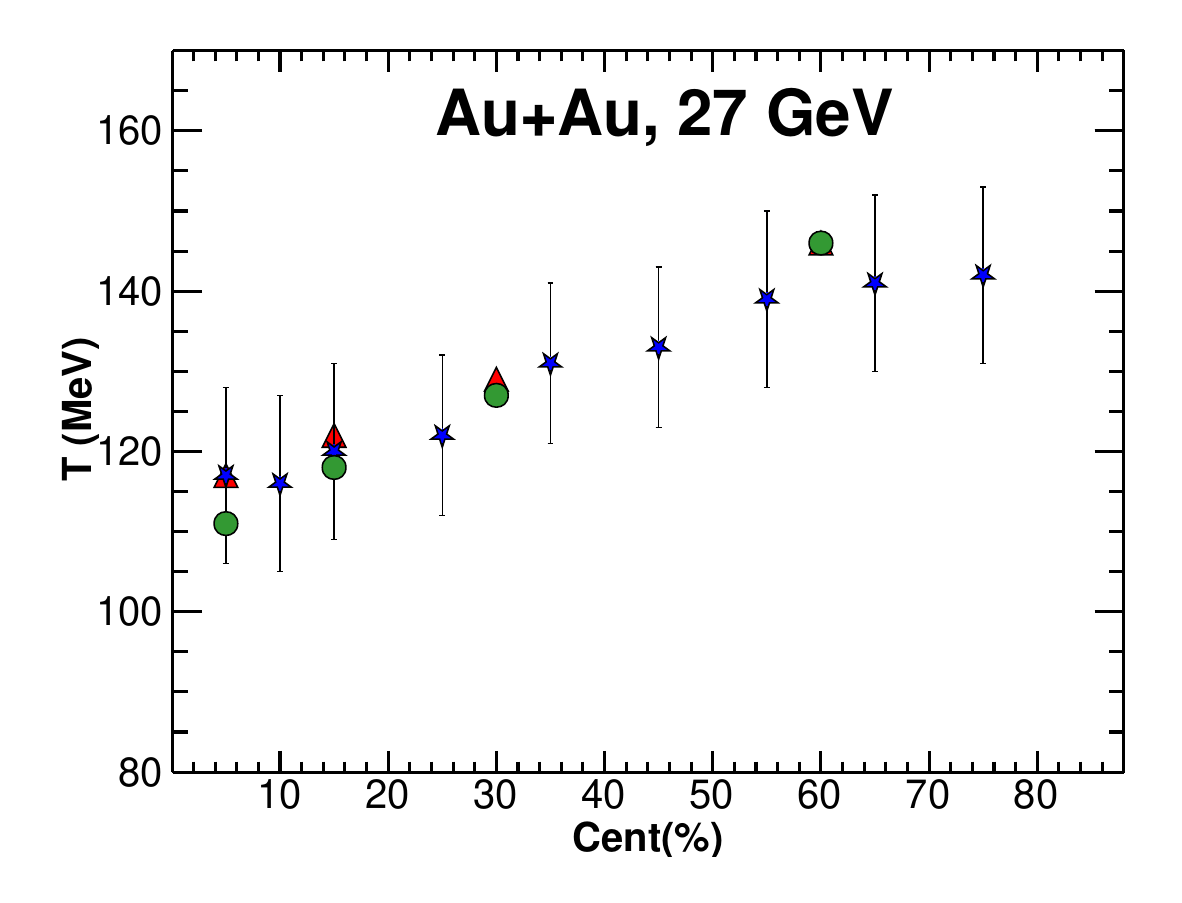}
        \hspace{-0.5cm}
        \includegraphics[width=0.335\linewidth, height=0.3\textwidth]{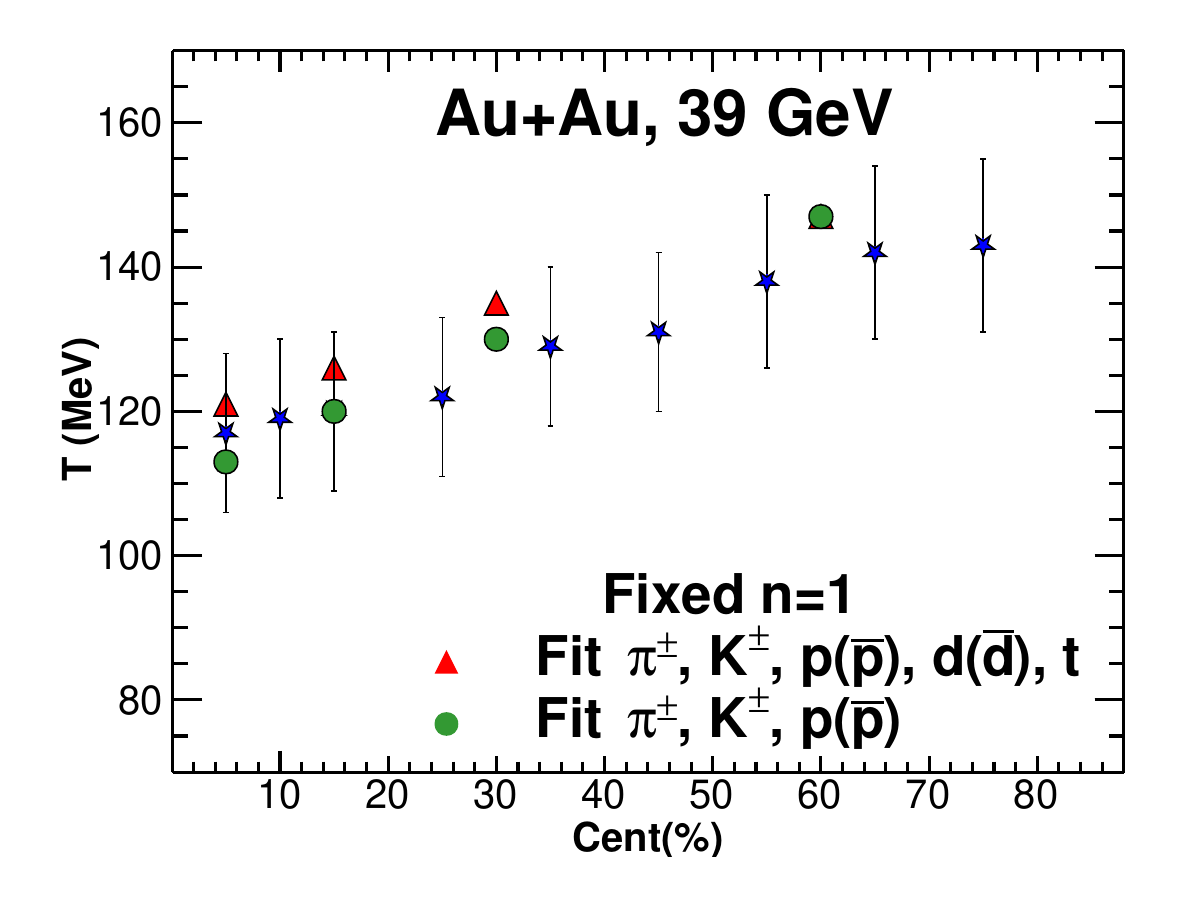}
        \hspace{-0.5cm}
        \includegraphics[width=0.335\linewidth, height=0.3\textwidth]{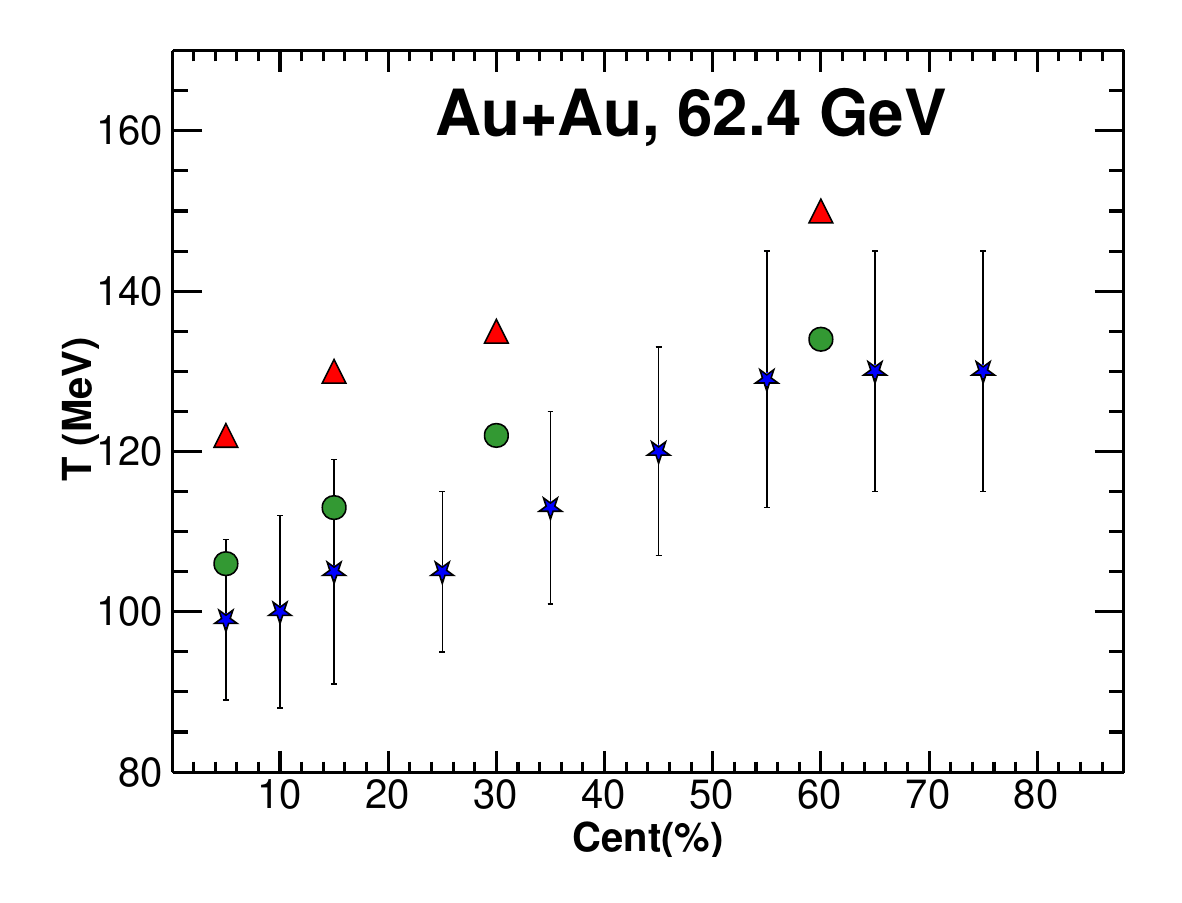}
        \hspace{-0.5cm}
        \includegraphics[width=0.335\linewidth, height=0.3\textwidth]{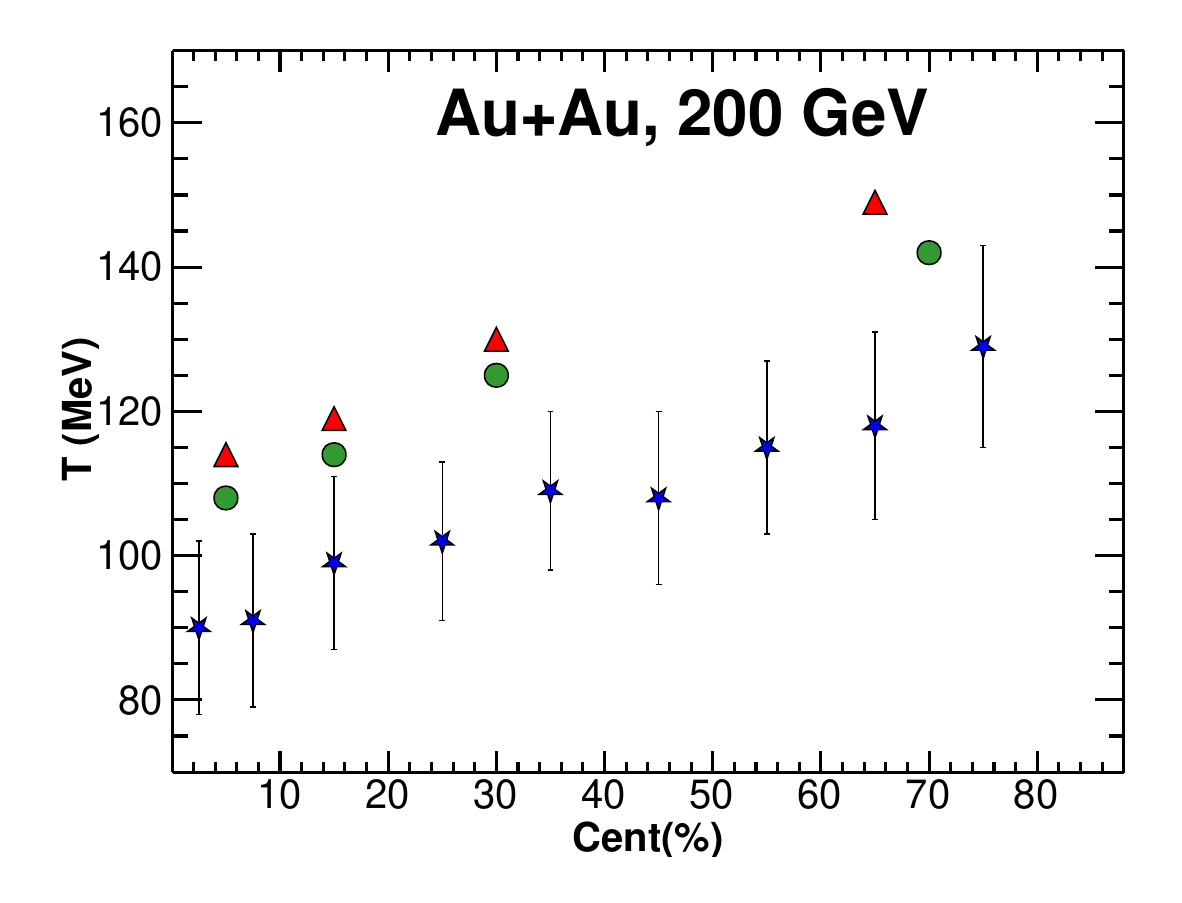}
        \hspace{-0.5cm}
        \includegraphics[width=0.335\linewidth, height=0.3\textwidth]{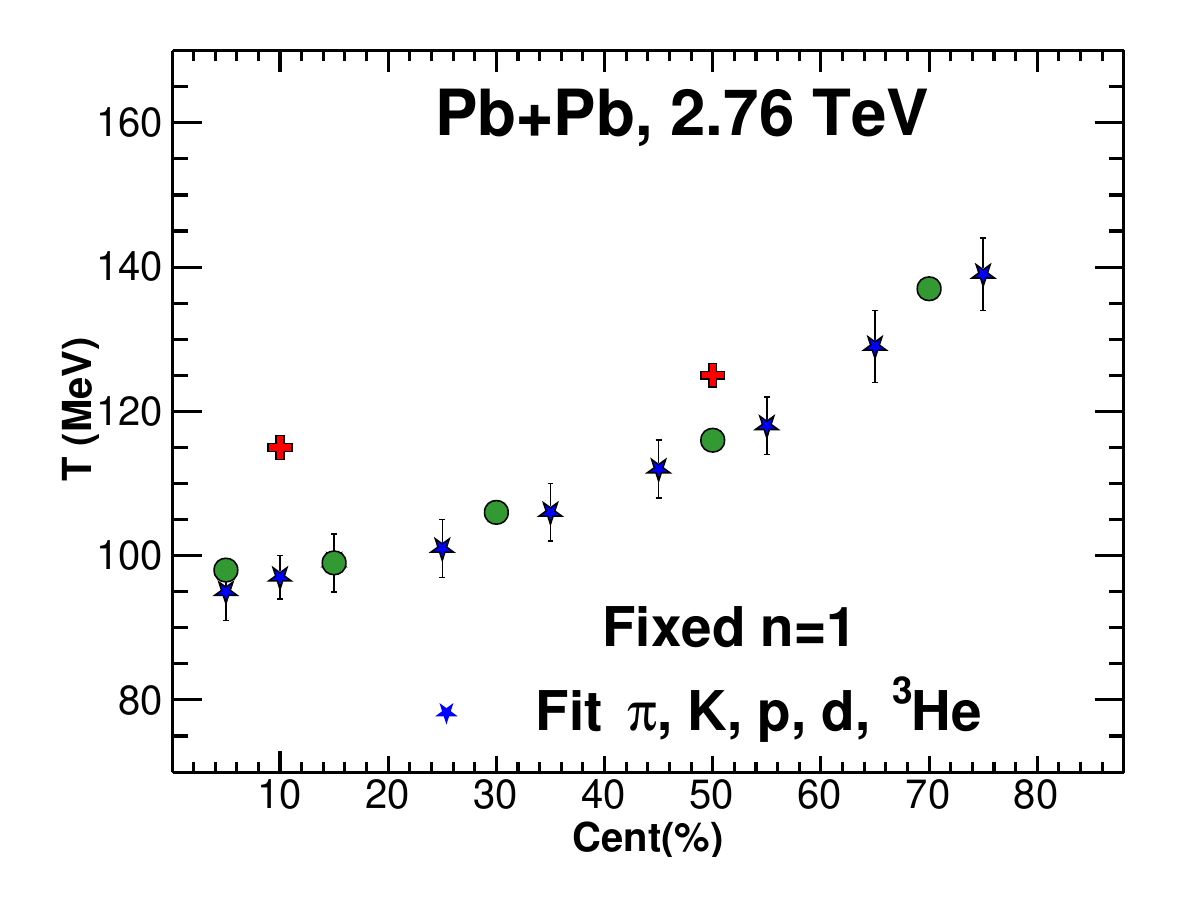}
    \end{subfigure}%
    \caption{(Color online) Centrality dependence of kinetic freeze-out temperature ($T_{kin}$) of light hadrons and light (anti-)nuclei in {\auau} collisions at {\sqrtsNN} = 7.7--200~GeV and {\pbpb} collisions at {\sqrtsNN} = 2.76 TeV. For {\pbpb} data, only bin-by-bin systematic uncertainties from Ref.~\cite{56} are considered. }\label{fig3}
\end{figure*}


\begin{figure*}
    \centering
    \begin{subfigure}
        \centering
        \includegraphics[width=0.335\linewidth, height=0.3\textwidth]{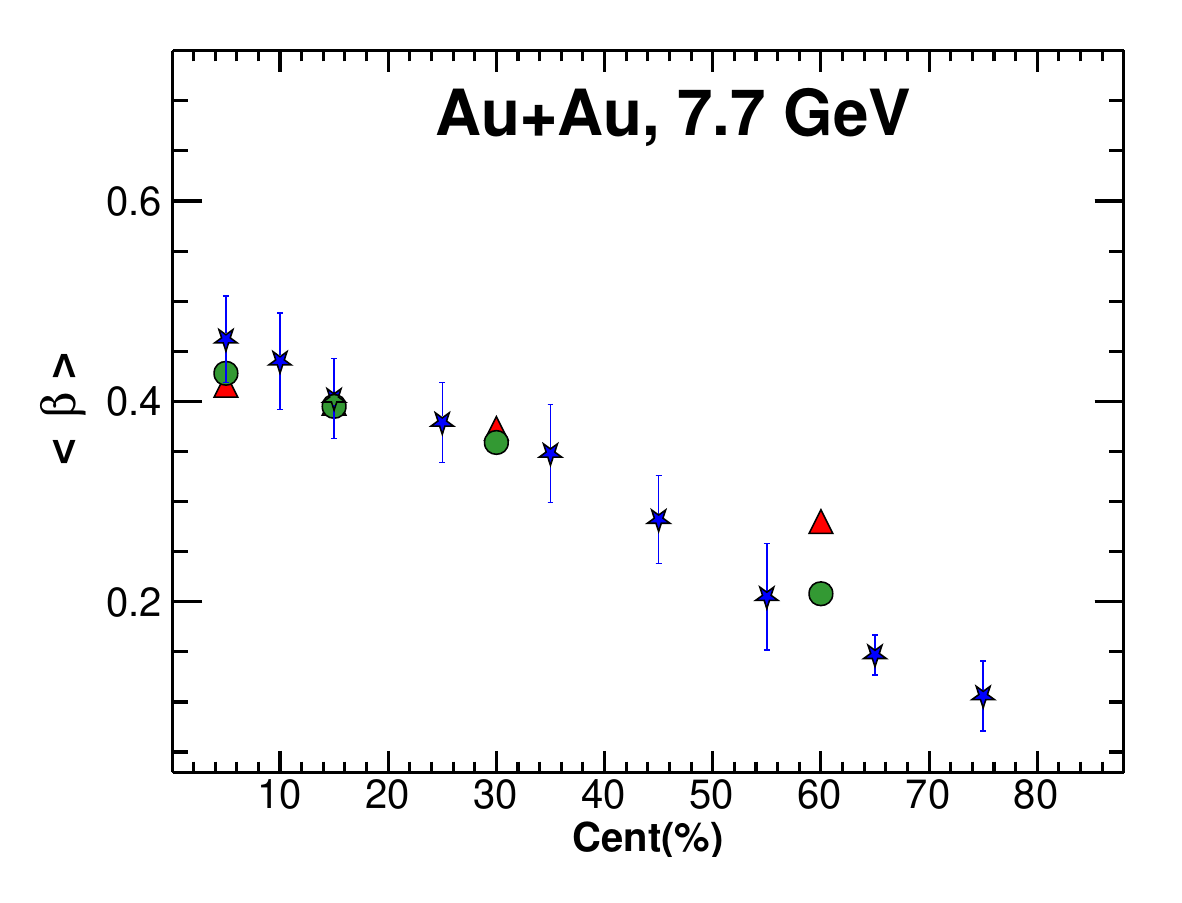} 
        \hspace{-0.5cm}
        \vspace{-0.5cm}
        \includegraphics[width=0.335\linewidth, height=0.3\textwidth]{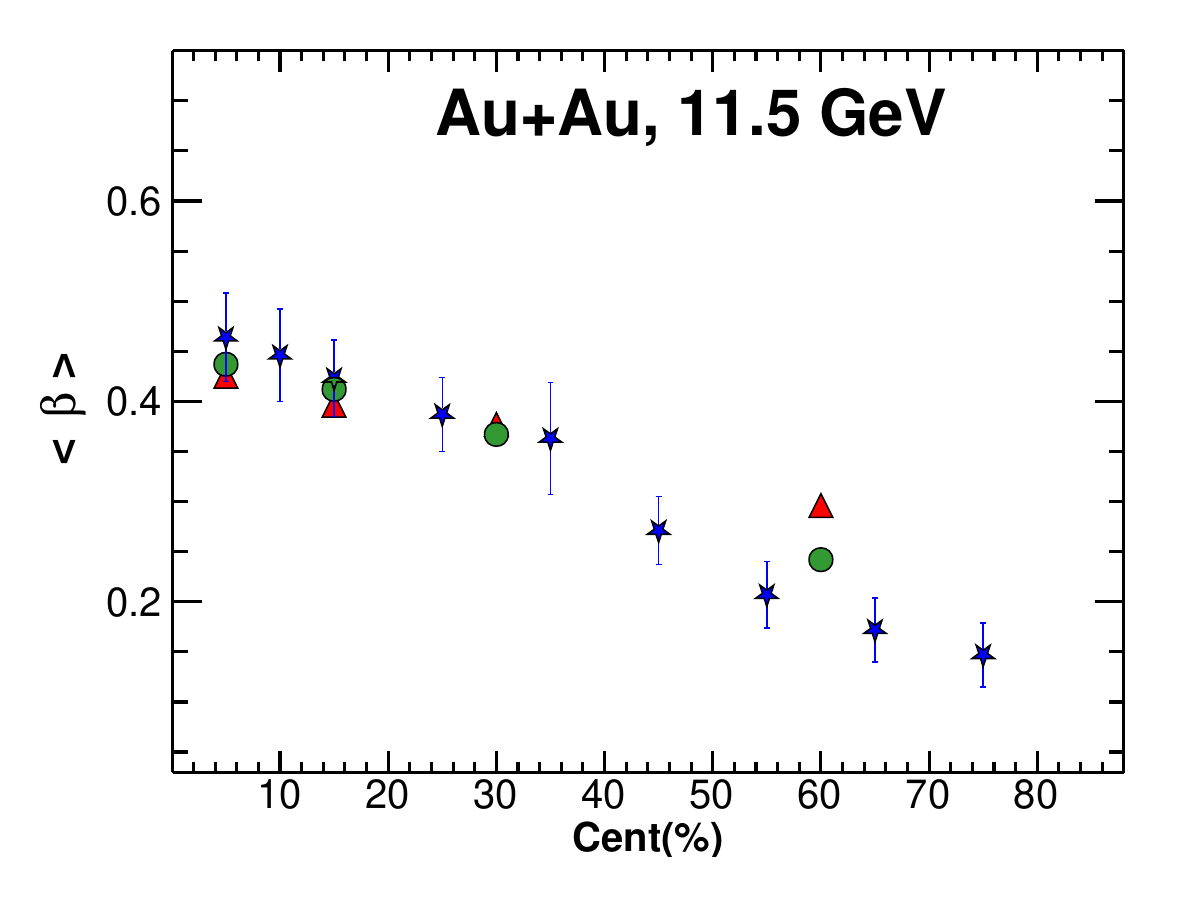}
        \hspace{-0.5cm}
        \includegraphics[width=0.335\linewidth, height=0.3\textwidth]{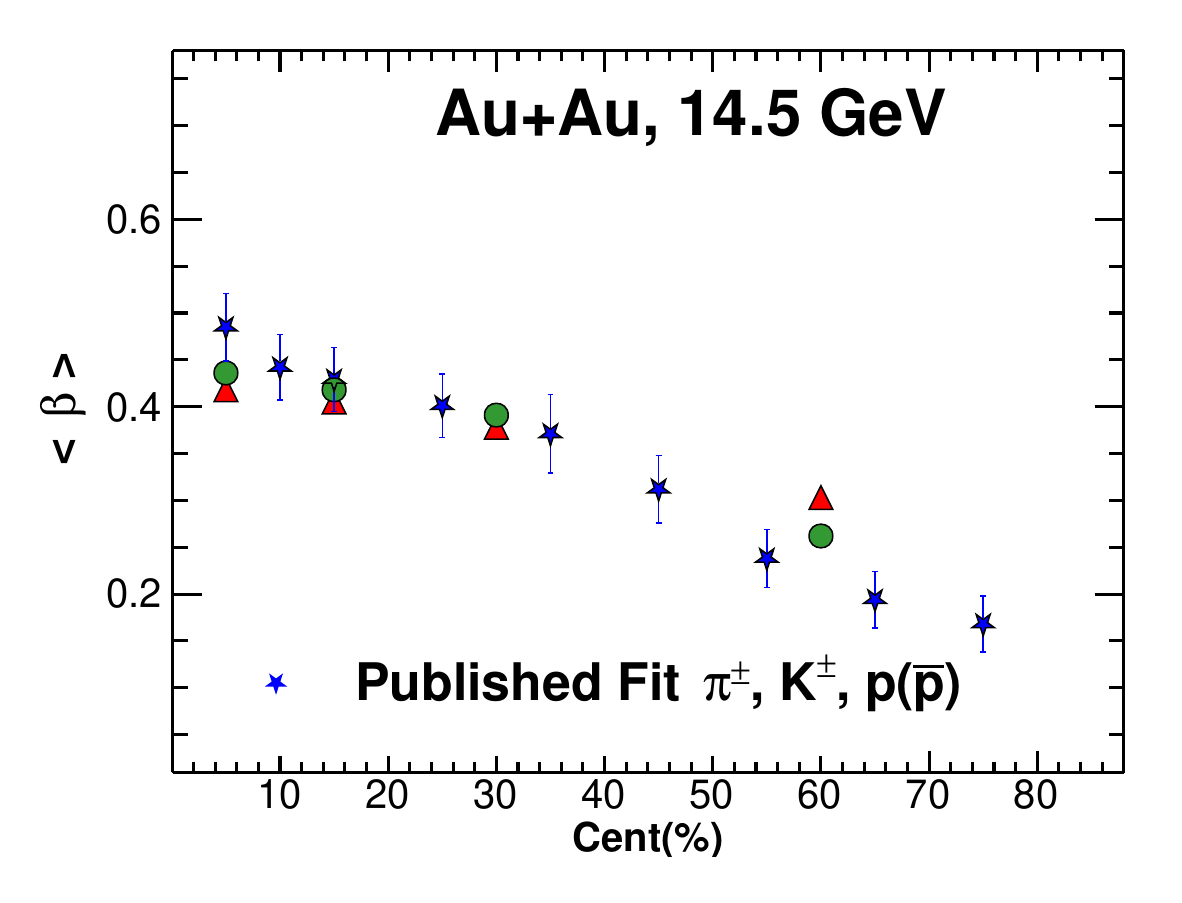}
        \hspace{-0.5cm}
        \vspace{-0.3cm}
        \includegraphics[width=0.335\linewidth, height=0.3\textwidth]{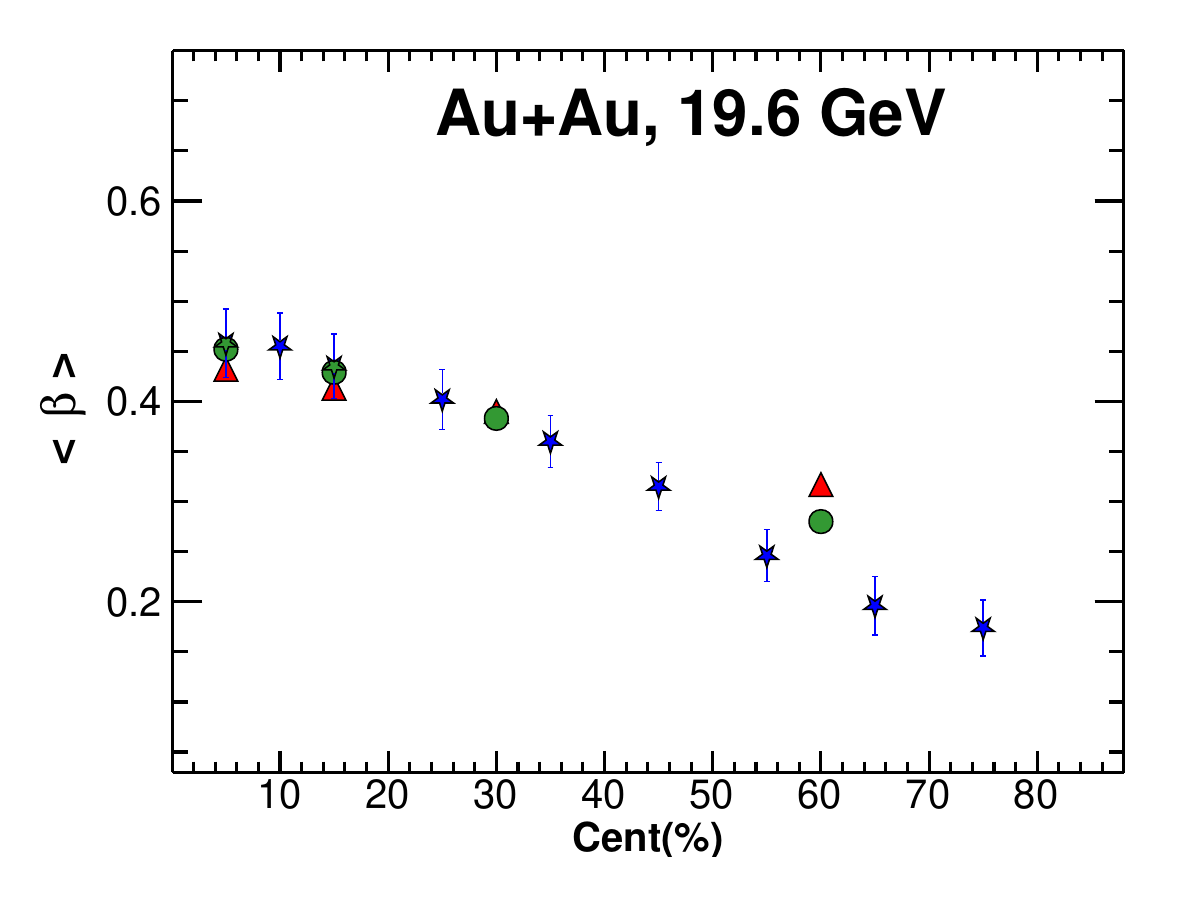}
        \hspace{-0.5cm}
        \includegraphics[width=0.335\linewidth, height=0.3\textwidth]{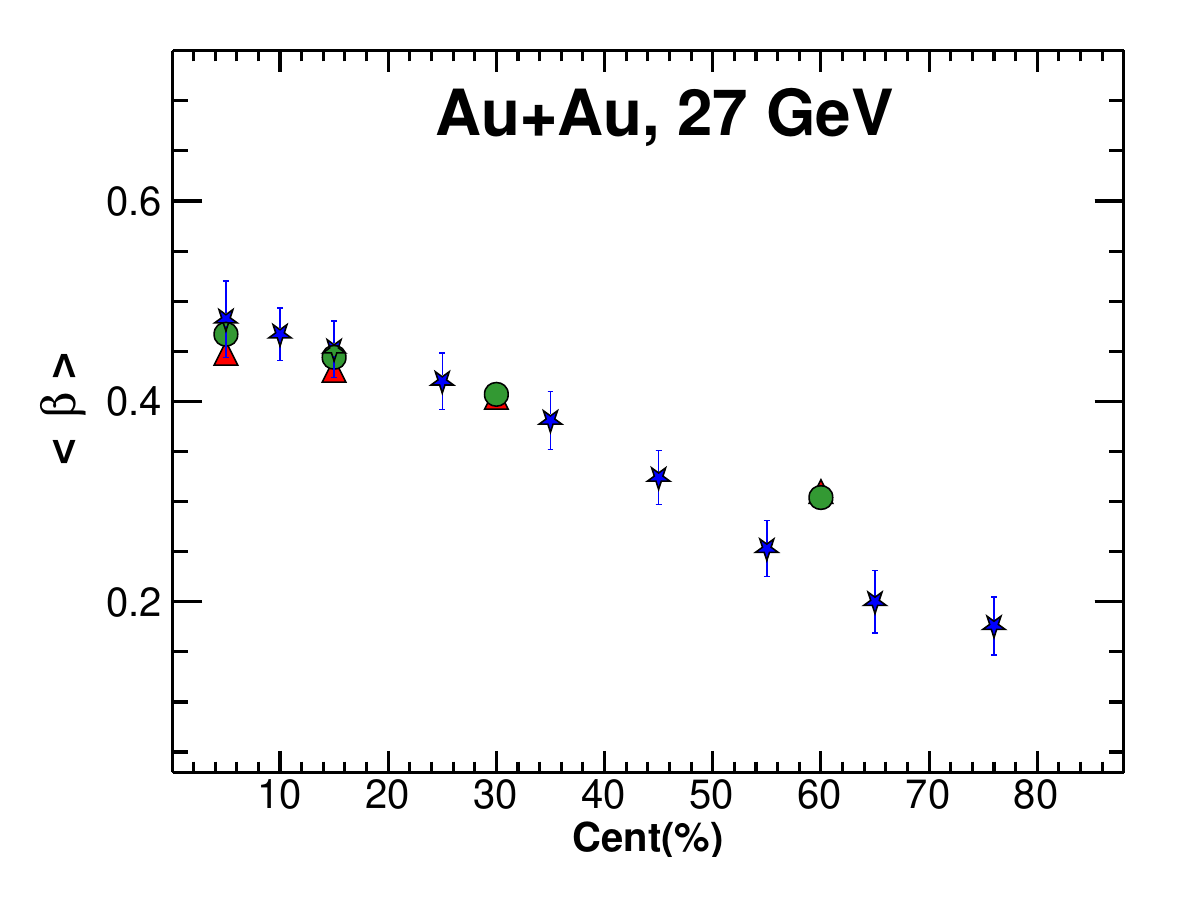}
        \hspace{-0.5cm}
        \includegraphics[width=0.335\linewidth, height=0.3\textwidth]{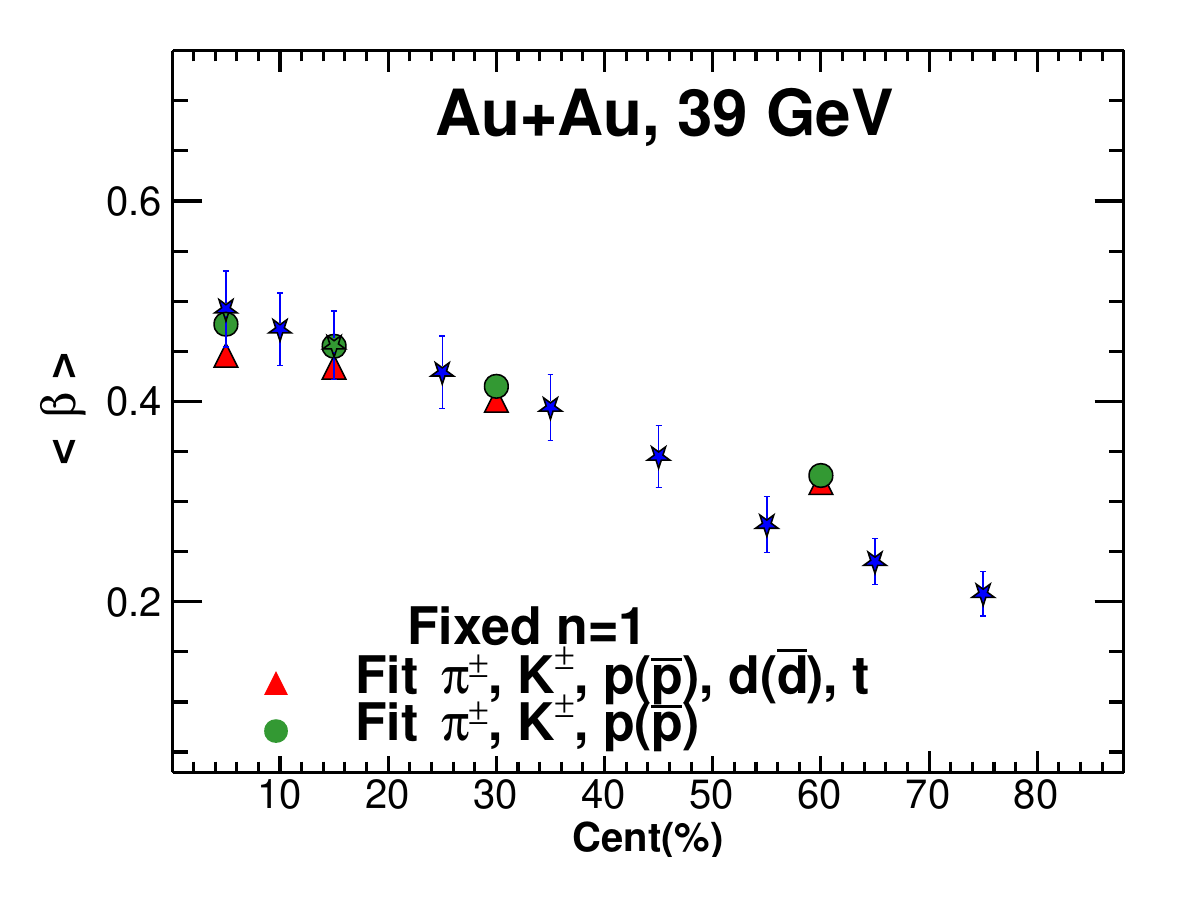}
        \hspace{-0.5cm}
        \includegraphics[width=0.335\linewidth, height=0.3\textwidth]{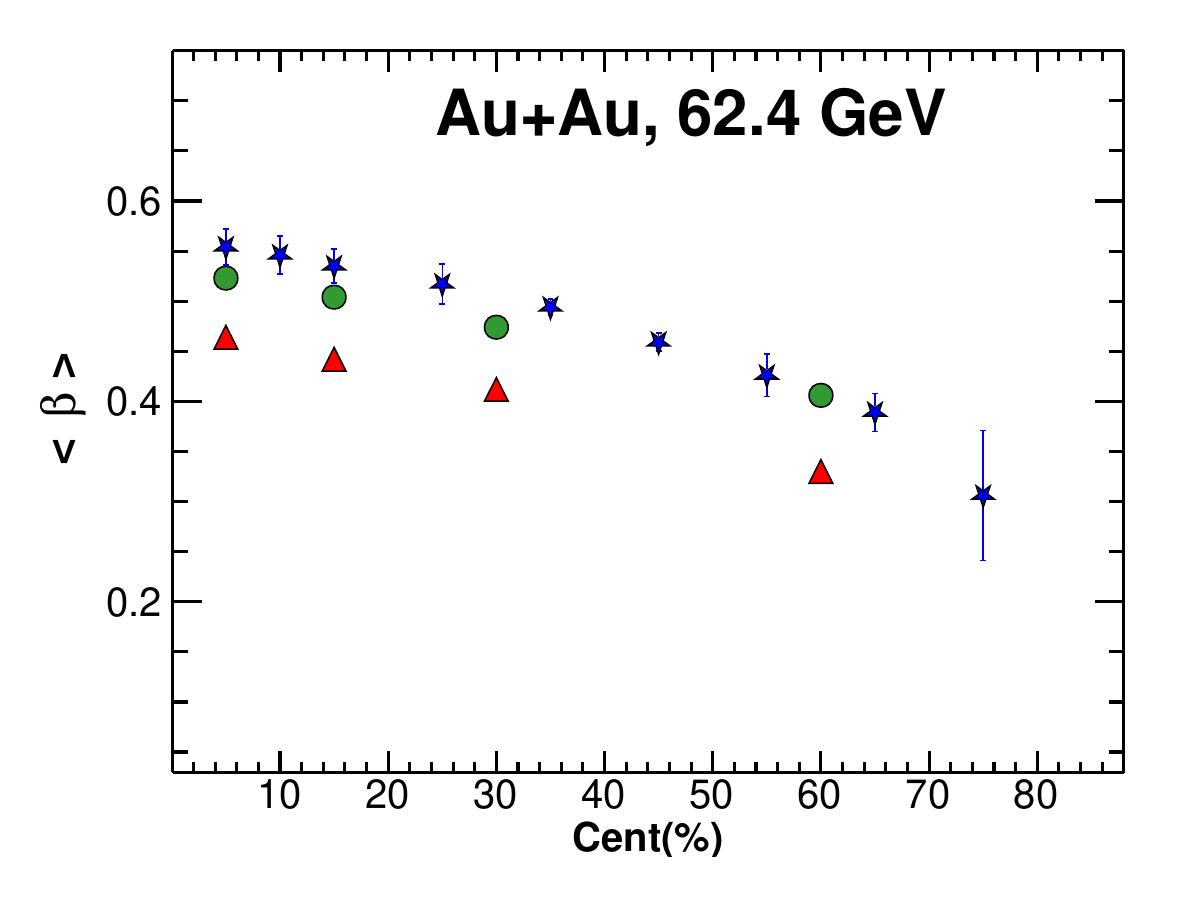}
        \hspace{-0.5cm}
        \includegraphics[width=0.335\linewidth, height=0.3\textwidth]{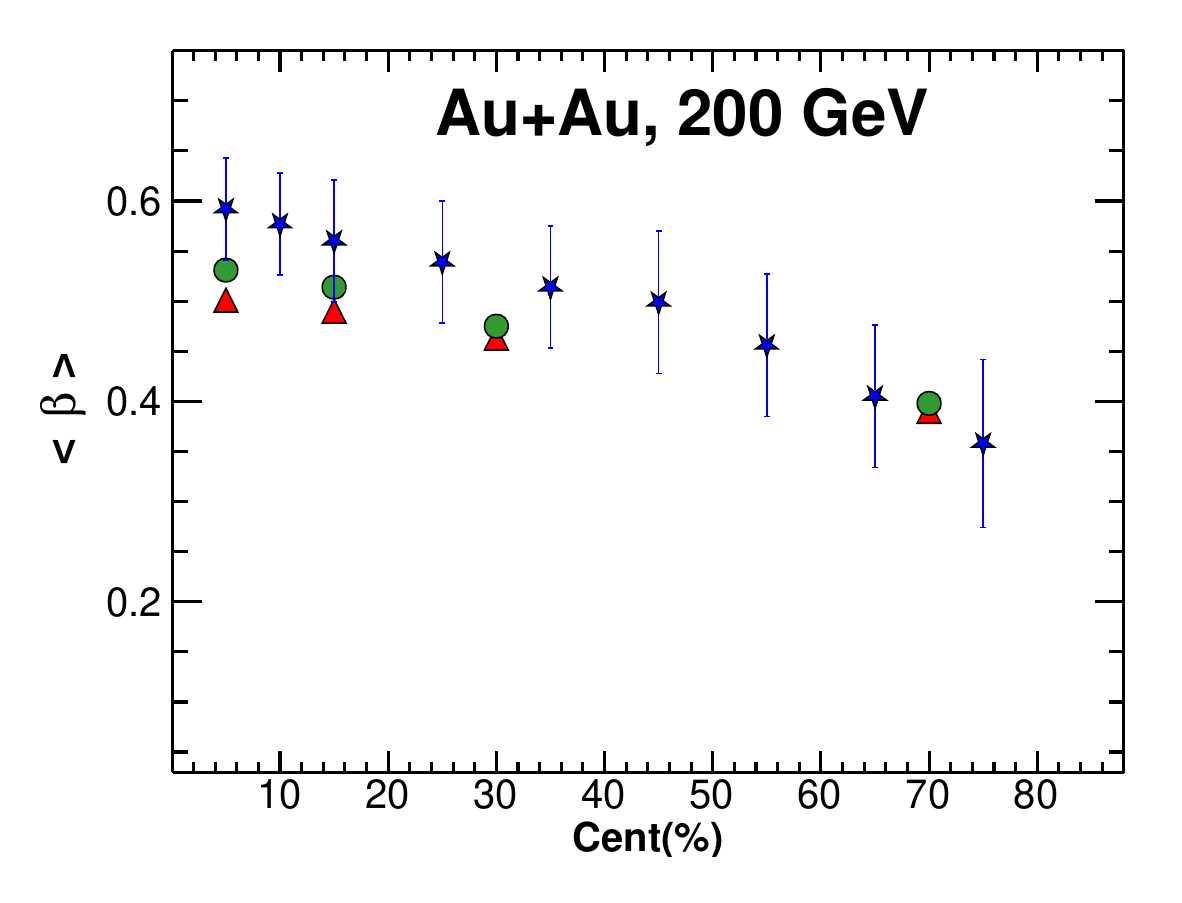}
        \hspace{-0.5cm}
        \includegraphics[width=0.335\linewidth, height=0.3\textwidth]{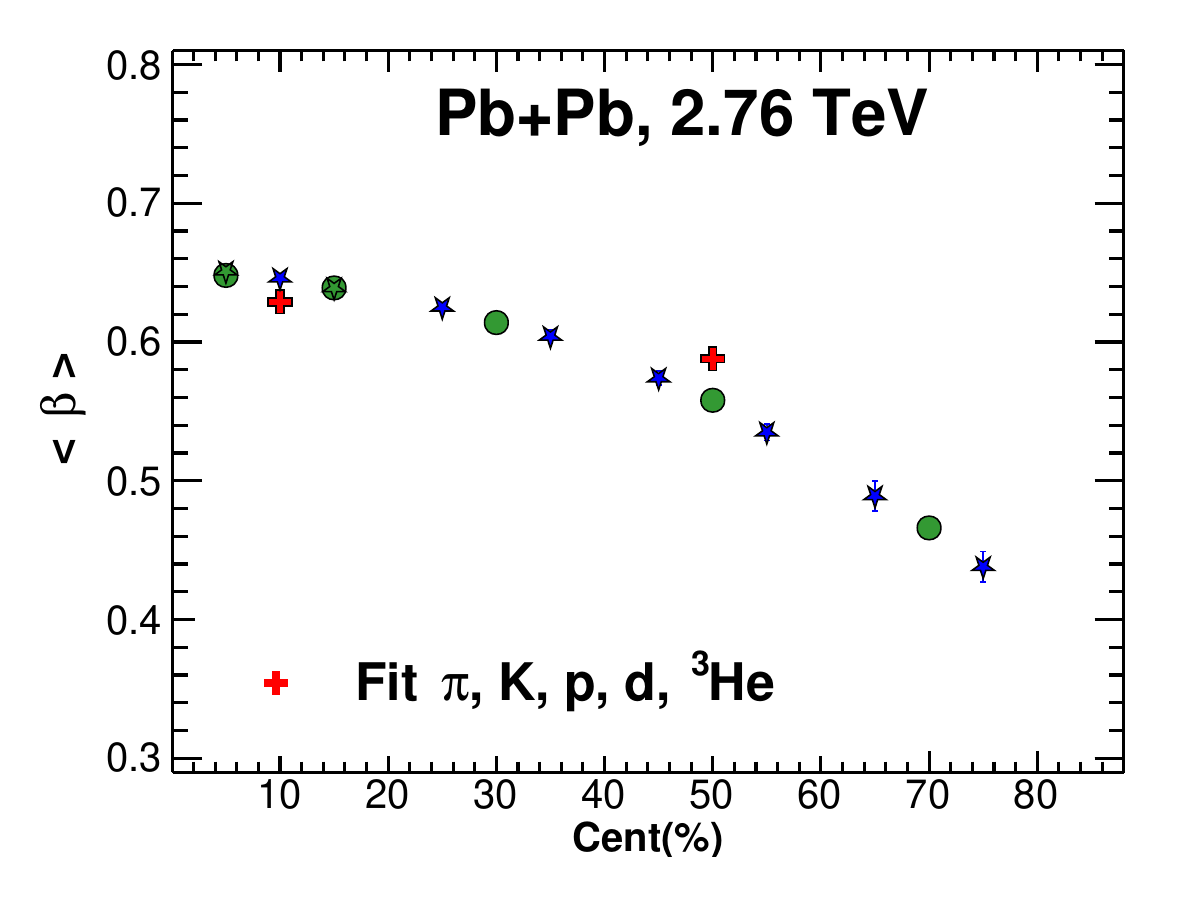}
    \end{subfigure}%
    \caption{(Color online) Centrality dependence of radial flow velocity ($\langle \beta \rangle$) of light hadrons and light (anti-)nuclei in {\auau} collisions at {\sqrtsNN} = 7.7 -- 200~GeV and {\pbpb} collisions at {\sqrtsNN} = 2.76~TeV. For {\pbpb} data, only bin-by-bin systematic uncertainties from Ref.~\cite{56} are considered.}\label{fig4}
\end{figure*}


Figures~\ref{fig3} and ~\ref{fig4} show the centrality dependence of temperature ($T_{kin}$) and radial flow velocity ($\langle \beta \rangle$), respectively. Parameters are extracted by simultaneous fits to the {\ppt} spectra of light hadrons along with light (anti-)nuclei in {\auau} collisions at {\sqrtsNN} = 7.7 -- 200~GeV from RHIC~\cite{55} and {\pbpb} collisions at {\sqrtsNN} = 2.76~TeV~\cite{56} from LHC. For simplicity, the centrality classes of $\pi^\pm$, $K^\pm$, $p(\bar p)$ are merged and refitted to make it consistent with those used for $d (\bar d)$ and $t (\bar t)$ measurements, i.e. 0--10\%, 10--20\%, 20--40\% and 40--80\%. Due to unavailability of the triton {\ppt} spectra from {\pbpb} collisions at {\sqrtsNN} = 2.76~TeV, we included ${}^{3}He$ in the fits instead. Blue markers represent the published results for $\pi^\pm$, $K^\pm$, and $p(\bar p)$. Green markers depict $\pi^\pm$, $K^\pm$, and $p(\bar p)$ data with combined centrality classes. Red markers correspond to $T_{kin}$ of $\pi^\pm$, $K^\pm$, $p(\bar p)$, and $d(\bar d)$. It is evident from Fig.~\ref{fig3} that $T_{kin}$ exhibits an increasing trend as a function of centrality for all collision energies. Furthermore, the inclusion of heavier particle species, such as light (anti-)nuclei, results in slightly higher temperatures compared to those of light hadrons. This effect is more prominent at energies {\sqrtsNN} $\geq$ 62.4~GeV. Fig.~\ref{fig4} shows that fitting both light hadrons with light (anti-)nuclei together leads to a smaller values of $\langle \beta \rangle$  as compared to those when only light hadrons are considered in the fits. To study the influence of light (anti-)nuclei data on the global characteristics, we repeat the fit by simultaneously excluding $d (\bar d)$ and $t$. This exhibits similar trend, but with slightly lower value of $T_{kin}$ and slightly larger values of $\langle \beta \rangle$. Other effects, like elliptic flow ($v_2$) of deutrons can also be used study the distinction between coalescence and statistical production. More details are discussed in Ref.~\cite{77}.  Moreover, the $T_{kin}$ in mid-central 20--40\% and peripheral collisions 40--80\% at energies {\sqrtsNN} $\leq$ 19.6~GeV is relatively lower as compared to central collisions and start to increase at energies at energies {\sqrtsNN} $\geq$ 27~GeV. On the other hand, $\langle \beta \rangle$ is slightly larger at the same centrality classes and energies. This may be related to the change in the collision geometry and different production mechanisms of light (anti-)nuclei at lower energies specifically in peripheral collisions where spectators plays an important role. Thus, a clear hierarchical structure seems to be seen in kinetic freeze-out parameters. We have also repeated the analysis by treating $n$ as a free parameter. It is shown that the best fit values of the extracted parameters $T_{kin}$ and $\langle \beta \rangle$ are found to be nearly unaltered when letting the parameter $n$ free for all energies. 

\begin{table*}[hbt!] 
\scriptsize{
\caption{Combined parameters extracted from the blast-wave fit to the {\ppt} spectra of different set of particle species (light hadrons and light (anti-)nuclei) in central (0--10\%) {\auau} collisions at {\sqrtsNN} = 27~GeV and {\pbpb} collisions at 2.76 TeV.  }
\vspace{-.50cm}
\begin{center}
\begin{tabular}{p{2cm}p{2cm}p{4cm}p{3cm}p{2cm}p{2cm}p{2cm}p{2cm}}\\ \hline\hline
     \sqrtsNN&Sets  &   Particle  & $<\beta> $ &    $T$ (MeV) & $n$ & $\chi^2$/ dof  \\\hline \hline
    
    27GeV &Set A  & $\pi^{\pm}$, $K^{\pm}$, $p$, $\bar{p}$, $d (\bar d), t$   & $0.465\pm0.003$ & $116\pm 3$ &$0.857\pm0.040$&0.53 \\
    &Set B  & $p(\bar p)$, $d(\bar d), t$   &  $0.464\pm0.017$ & $129\pm 9$ & $0.723\pm0.104$&0.72  \\
    &Set C & $d(\bar d), t$   & $0.460\pm0.017$ & $121\pm 10$ & $0.840\pm0.132$&1.25  \\
    &Set D & $\pi^{\pm}$, $K^{\pm}$, $p$, $\bar{p}$   & $0.475\pm0.033$ & $116\pm 3$ &$0.672\pm 0.198$ &0.28  \\
    
    \hline

    2.76 TeV &Set A  & $\pi^{\pm}$, $K^{\pm}$, $p$, $\bar{p}$, $d , ^3$He   & $0.673\pm0.003$ & $115\pm 3$ &$0.698\pm0.015$&0.75 \\
    &Set B  & $p(\bar p)$, $d(\bar d), ^3$He   &  $0.633\pm0.08$ & $149\pm 7$ & $0.556\pm0.031$&0.41  \\
    &Set B$^\prime$   & $p(\bar p)$, $d(\bar d), ^3$He   &  $0.673\pm0.003$ & $132\pm 5$ & $0.688\pm0.025$&2.3  \\
    &Set C & $d, ^3$He  & $0.622\pm0.007$ & $80\pm 17$ & $0.779\pm0.043$&0.18  \\
    &Set D & $\pi^{\pm}$, $K^{\pm}$, $p$, $\bar{p}$   & $0.640\pm0.018$ & $98\pm5$ &$0.701\pm 0.120$ &0.25  \\
    
    \hline

\end{tabular}
\label{table2}
\end{center}} 
\end{table*}

In the coalescence model nuclei are assumed to be formed at a later stage of the collisions precisely at or after the kinetic freeze-out. Under this condition, one can anticipate that the {\ppt} spectra of both protons and (anti-)nuclei reflect a common temperature and velocity field of the source at or later stage of the nuclear cluster formation. To provide a more detailed explanation, we have categorized various particle species into distinct sets and performed the BW fit. Set A includes $\pi^\pm$, $K^\pm$, $p(\bar p)$, $d(\bar d)$ and $t$; Set B comprises $p(\bar p)$, $d(\bar d)$ and $t$; Set C consists of $d(\bar d)$ and $t$; and Set D contains only $\pi^\pm$, $K^\pm$, $p(\bar p)$. Note that we use ${}^{3}He$ instead of $t$ where no data for triton is available at LHC energies. The extracted parameters from different set of combinations in (0--10\%) central {\auau} collisions at {\sqrtsNN} = 27~GeV are listed in Table~\ref{table2} as an example. The similar trend is observed for the rest of energies. It is evident from the Table~\ref{table2} that including $p (\bar p)$ in the fit with light (anti-)nuclei exhibits a similar velocity profile as observed when $\pi^\pm$, $K^\pm$, $p(\bar p)$ are considered in the fit (Set A). However, there is a notable difference in the kinetic freeze-out temperature ($T_{kin}$) and is found to be significantly larger ($129\pm 9$)~MeV which is unexpected in the context of final-state coalescence and requires further investigation. The observed results aligns with the hypothesis of statistical hadronization in which light (anti-)nuclei are formed close to the phase boundary without undergoing significant interactions at later stages of the system evolution. The observed scenario might be explained by considering pre-hadronic multiquark states and has been discussed in detail in Ref.~\cite{36, 76}.  We repeated the exercise for {\pbpb} collisions at {\sqrtsNN} = 2.76 TeV and observed similar findings. The corresponding extracted parameters are listed in Table~\ref{table2}. The velocity profile of Set B, when including $p(\bar p)$ in the fit along with light (anti-)nuclei, appears to be relatively lower compared to Set A. However, $T_{kin}$ is significantly larger ($149 \pm 7$) MeV. To understand this, we introduced another set, denoted as Set B$^\prime$, in which we keep the same velocity profile as that of Set A. As a result, the $T_{kin}$ is noticeably reduced to ($132 \pm 5$) MeV and is consistent with ALICE experimental results from {\pbpb} collisions at {\sqrtsNN} = 5.02 TeV~\cite{76}.

For further investigation, a simultaneous BW fit is performed to the {\ppt} spectra of only light (anti-)nuclei in {\auau} collisions at {\sqrtsNN} = 27 GeV (Set C). The extracted $T_{kin}$ is $121\pm10$~MeV is in good agreement within uncertainties with the results of Set A, where all particles are included in the fit and are listed in Table~\ref{table2}. This finding appears to be more consistent with the expectations of the coalescence model. It suggests that protons freeze-out earlier at a higher temperature as indicated by Set B. Subsequently, these protons and neutrons coalesce to form nuclei at a later stage through the coalescence process. When examining Set D, which considers only 0--10\% central $\pi^\pm$, $K^\pm$, $p(\bar p)$ {\ppt} spectra in {\auau} collisions at {\sqrtsNN} = 27~GeV in the fit the resulting $T_{kin}$ is $116\pm 3$~MeV. On the other hand, in {\pbpb} collisions at {\sqrtsNN} = 2.76 TeV  the value of $T_{kin}$ from Set D relatively lower than that obtained from Set A, suggesting that the inclusion of the lightest particles contributes to the observation of lower kinetic freeze-out temperatures. It is observed that $T_{kin}$ in 0--10\% central {\auau} collisions at {\sqrtsNN} = 27~GeV is considerably higher than those in 0--10\% central {\pbpb} collisions reported at LHC energies, as documented in Ref.~\cite{56}. It is important to emphasize that lighter particles are more likely to be influenced by contributions originating from resonance decays and hard scatterings across a broader {\ppt} range as compared to those of heavier particles. It is important to mention that the BW model is only a simplified hydrodynamic approach to understand the collective motion and freeze-out of particles and has certain limitations; therefore, one should be cautious when making any strong conclusions.

\begin{figure}[h]
\centering
\includegraphics[width=0.45\textwidth]{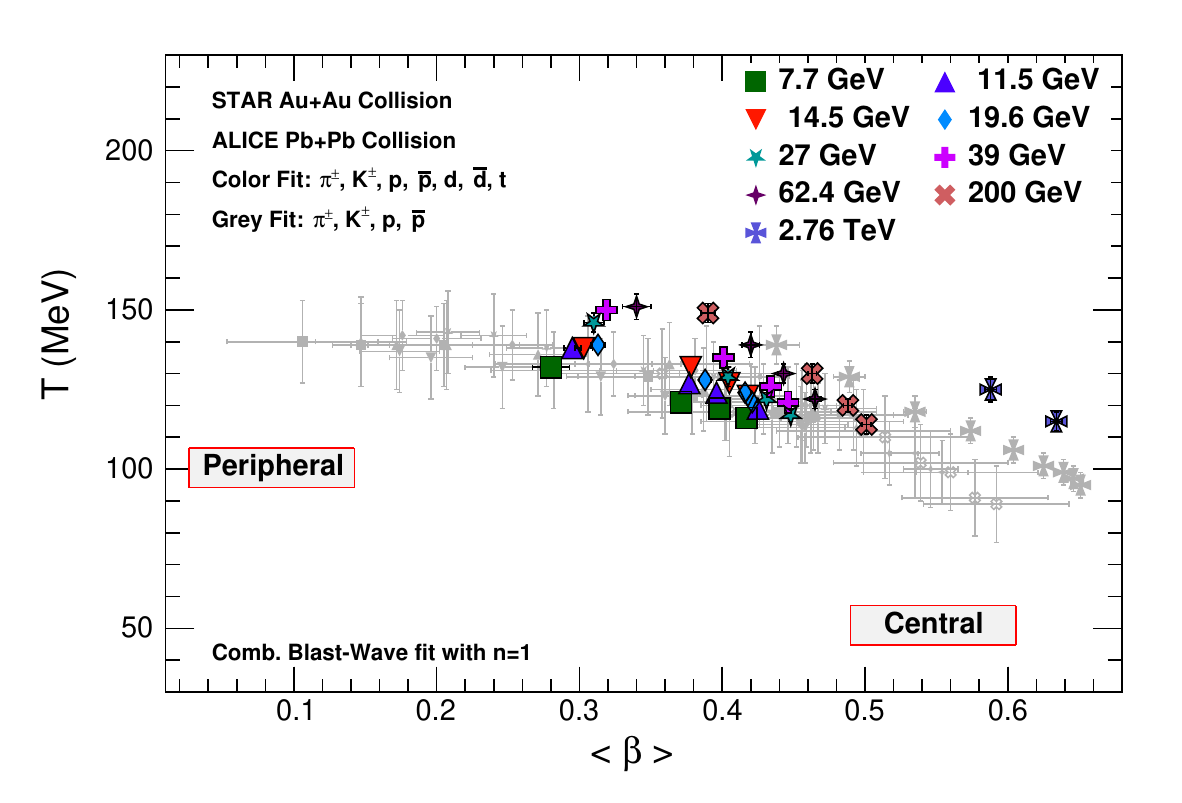}
\caption{(Color online) Variation of kinetic freeze-out temperature ($T_{kin}$) with radial flow velocity ($\langle \beta \rangle$) in {\auau} collisions at {\sqrtsNN} = 7.7 -- 200~GeV and {\pbpb} collisions at {\sqrtsNN} = 2.76~TeV for different centrality classes. The centrality increases from left to right for a given energy. }
\label{fig6}
\end{figure}

Figure~\ref{fig6} shows the variation of extracted $T_{kin}$ with $\langle \beta \rangle$ for different centrality classes and energies. The grey markers are the simultaneous fit to $\pi^\pm, K^\pm$ and $p(\bar  p)$ {\ppt} spectra only and are taken from Ref.~\cite{55}. The colored markers are our combined BW fit to the {\ppt} spectra of $\pi^\pm, K^\pm, p(\bar  p), d (\bar d), t$ and $ {}^{3}He$. We observe that including light (anti-)nuclei in the analysis leads to a slight increase in $T_{kin}$ and slight decrease in $\langle \beta \rangle$. This might be an indication of earlier freeze-out of the system by adding more heavier particle to the fit and might be more sensitive to the effects of radial flow. Moreover, the extracted parameters show strong two dimensional anti-correlation, i.e. higher values of $T_{kin}$ corresponds to lower values of $\langle \beta \rangle$ and vice-versa.

\begin{figure}[!ht]
\centering
\includegraphics[width=0.45\textwidth]{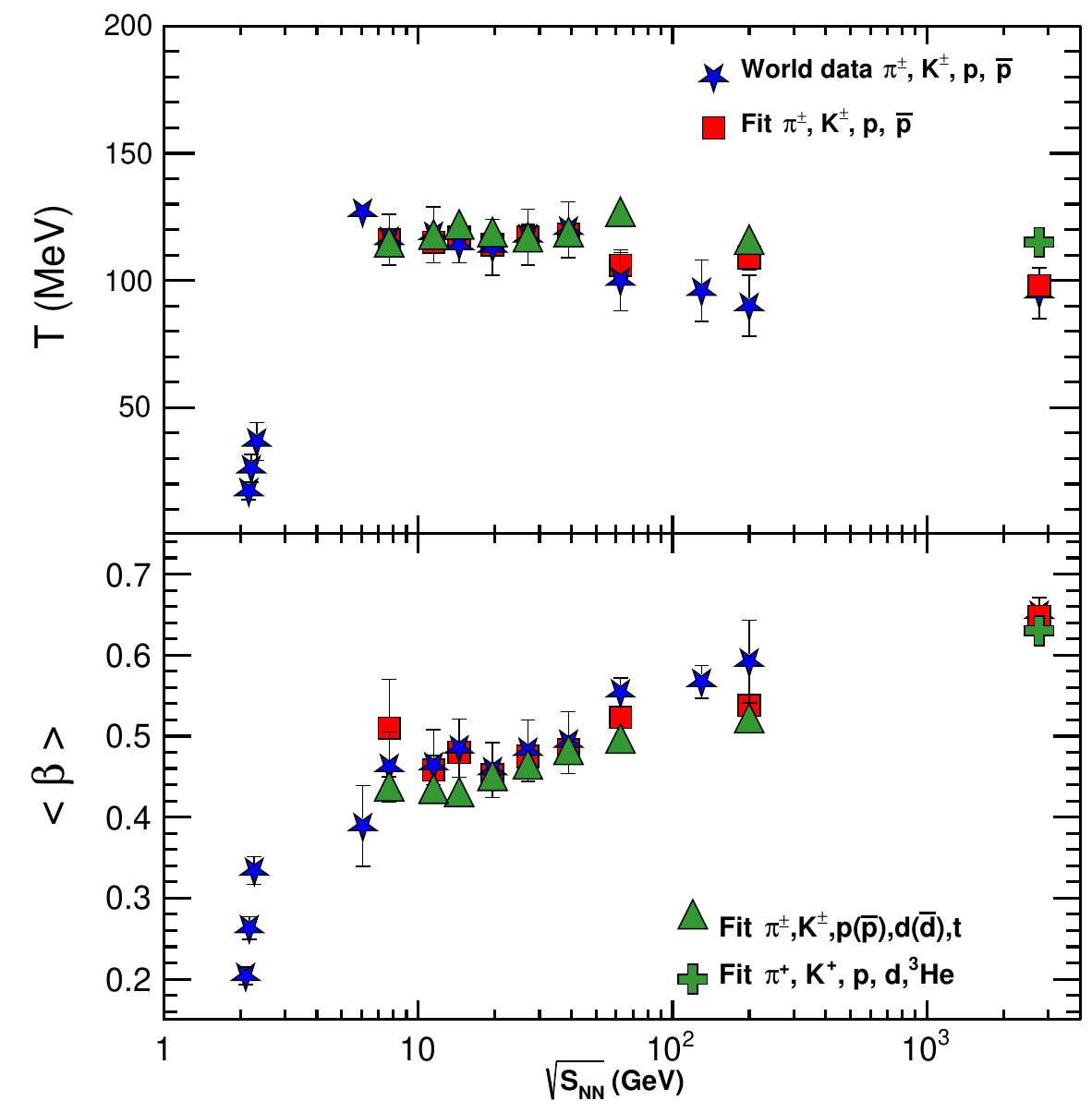}

\caption{(Color online) Energy dependence of kinetic freeze-out temperature ($T_{kin}$) (top) and radial flow velocity ($\langle \beta \rangle$) (bottom) for central heavy-ion collisions.}
\label{fig7}
\end{figure}

Energy dependence of kinetic freeze-out temperature ($T_{kin}$) and radial flow velocity ($\langle \beta \rangle$) extracted from the fit to {\ppt} spectra of light hadrons along with light (anti-)nuclei is shown in Fig.~\ref{fig7}. The red markers represent our fit to the {\ppt} spectra of $\pi^\pm, K^\pm, p(\bar  p)$ in central (0--10\%) {\auau} collisions at {\sqrtsNN} = 7.7 -- 200~GeV and {\pbpb} collisions at {\sqrtsNN} = 2.76~TeV. The blue markers are the world published data for identified particles. Green markers correspond to the BW fit to the {\ppt} spectra of light hadrons together with light (anti-)nuclei. It is observed that $T_{kin}$ is relatively higher and $\langle \beta \rangle$ is lower when we include light (anti-)nuclei in the fit. Overall, they follow the trend of the world data. $T_{kin}$ remains almost constant from energies {\sqrtsNN} = 7.7 -- 39~GeV and starts decreasing towards LHC energies. On the other hand, the transverse flow velocity $\langle \beta \rangle$ increases over the entire energy range.

\section{Conclusions}
\label{Conclusion}

The analysis of transverse momentum {\ppt} spectra of produced particles provide insights into the particle production mechanisms as well as the freeze-out conditions of the system. We reported the comprehensive investigation of the kinetic freeze-out properties of identified hadrons ($\pi^\pm$, $K^\pm$ and $p(\bar p)$) along with light (anti-)nuclei $d (\bar d)$, $t (\bar t)$ and ${}^{3}He$ in relativistic heavy-ion collisions at RHIC and LHC energies. Blast-Wave analysis of the {\ppt} distribution of identified hadrons together with light (anti-)nuclei results in a slight increase in $T_{kin}$ and slight decrease in $\langle \beta \rangle$ in comparison to those when only identified hadrons are included in the fit. This suggests that relatively heavier particles like light (anti-)nuclei participate in a common flow field. Including only $p(\overline{p})$ in the fit together with light (anti-)nuclei leads to a surprisingly large extracted $T_{kin}$. The strong anti-correlation between $T_{kin}$ and $\langle \beta \rangle$, i.e., the value of $T_{kin}$ increases from central to peripheral collisions, indicating a shorter-lived fireball in peripheral collisions. Conversely, the value of $\langle \beta \rangle$ decreases from central to peripheral collisions, consistent with the picture of a more rapid expansion of the system in central collisions at both RHIC and LHC energies. At RHIC energies, the extracted $T_{kin}$ exhibits weak energy dependence. However, it decreases at LHC energies, likely due to the longer scattering phase at higher energies. On the other hand, $\langle \beta \rangle$ shows an increasing trend from RHIC to LHC energies. Further investigations into other important aspects, such as flow, can be used to distinguish between different production mechanisms of light (anti-)nuclei.

\section{Acknowledgement}
The authors would like to acknowledge M. Aamir Shahzad, A. M. Khan and M. Uzair. Aslam for useful discussions.

\bibliography{bib.bib}


\section{Appendix}
\label{Appen}

\begin{table*}[!ht] 
\scriptsize{
\caption{Extracted kinetic freeze-out parameters in {\auau} collisions at {\sqrtsNN} = 7.7 - 39 GeV from RHIC and {\pbpb} collisions at {\sqrtsNN} = 2.76 TeV with fixed $n = 1$ quoted with errors.}
\vspace{-.50cm}
\begin{center}
\begin{tabular}{p{2.5cm}p{2cm}p{2cm}p{2cm}p{2cm}p{2cm}}\\ \hline\hline
     Collision & Centrality &   $<\beta> $ &    $T$ (MeV) & $\chi^2$/ dof  \\\hline
    {\auau} 7.7 GeV   & $0-10\%$ & $0.416\pm0.004$   & $117\pm2$        & 2.40\\
            & $10-20\%$ &$0.398\pm0.004$   & $119\pm2$         & 0.77  \\
            & $20-40\%$ &$0.373\pm0.004$   & $121\pm2$        & 1.30\\
            & $40-80\%$ &$0.280\pm0.013$   & $131\pm3$         &  3.20\\
            \hline
    {\auau} 11.5 GeV   & $0-10\%$ & $0.425\pm0.003$   & $119\pm2$        & 0.61\\
            & $10-20\%$ &$0.396\pm0.004$   & $124\pm2$         & 3.10  \\
            & $20-40\%$ &$0.377\pm0.004$   & $127\pm2$        & 0.80\\
            & $40-80\%$ &$0.435\pm0.04$   & $118\pm2$        & 2.11 \\
  \hline
 {\auau} 14.5 GeV   & $0-10\%$ & $0.418\pm0.001$   & $123\pm2$    & 0.44\\
            & $10-20\%$ &$0.405\pm0.004$   & $127\pm2$& 0.30\\
            & $20-40\%$ &$0.378\pm0.004$   & $132\pm2$ & 0.33\\
            & $40-80\%$ &$0.303\pm0.006$   & $138\pm2$         &1.84  \\

  \hline
   {\auau} 19.6 GeV   & $0--10\%$ & $0.432\pm0.004$   & $125\pm2$        & 0.69\\
            & $10-20\%$ &$0.413\pm0.004$   & $125\pm2$   &0.34   \\
            & $20-40\%$ &$0.390\pm0.004$   & $128\pm2$& 0.55\\
            & $40-80\%$ &$0.317\pm0.006$   & $137\pm2$& 0.60 \\

  \hline
  {\auau} 27 GeV   & $0-10\%$ & $0.448\pm0.003$   & $117\pm3$        & 0.64\\
            & $10-20\%$ &$0.431\pm0.003$   & $122\pm3$         & 0.43  \\
            & $20-40\%$ &$0.404\pm0.004$   & $129\pm3$      & 0.31\\
            & $40-80\%$ &$0.310\pm0.007$   & $146\pm3$         & 0.60 \\
            
  \hline
   {\auau} 39 GeV   & $0-10\%$ & $0.446\pm0.003$   & $121\pm3$       & 1.29\\
            & $10-20\%$ &$0.434\pm0.003$   & $126\pm3$          & 0.45  \\
            & $20-40\%$ &$0.401\pm0.004$   & $135\pm3$         & 0.39\\
            & $40-80\%$ &$0.319\pm0.005$   & $150\pm3$           & 0.72\\
            \hline
     {\auau} 62.4 GeV   & $0-10\%$ & $0.464\pm0.004$   & $122\pm3$ & 1.43\\
            & $10-20\%$ &$0.442\pm0.004$   & $130\pm3$ & 1.32  \\
            & $20-40\%$ &$0.412\pm0.006$   & $139\pm4$ & 1.01\\
            & $40-80\%$ &$0.317\pm0.010$   & $150\pm4$ & 0.85\\
            \hline
     {\auau} 200 GeV   & $0-10\%$ & $0.501\pm0.003$   & $114\pm3$ & 1.01\\
            & $10-20\%$ &$0.488\pm0.003$   & $120\pm3$ &0.86  \\
            & $20-40\%$ &$0.463\pm0.003$   & $130\pm4$ & 0.58\\
            & $40-80\%$ &$0.390\pm0.005$   & $149\pm4$ & 1.90\\
            \hline
     {\pbpb} 2760 GeV   & $0-20\%$ & $0.630\pm0.004$   & $123\pm4$       & 2.90\\
            & $20-80\%$ &$0.585\pm0.004$   & $125\pm5$ & 1.66  \\
           
    \hline
 \hline
\end{tabular}
\label{tab2}
\end{center}} 
\end{table*}

\begin{table*}[h] 
\scriptsize{
\caption{Extracted kinetic freeze-out parameters in {\auau} collisions at {\sqrtsNN} = 7.7 - 39 GeV from RHIC and {\pbpb} collisions at {\sqrtsNN} = 2.76 TeV with free $n$ quoted with errors.}
\vspace{-.50cm}
\begin{center}
\begin{tabular}{p{2.5cm}p{2cm}p{2cm}p{2cm}p{2cm}p{2cm}p{2cm}}\\ \hline\hline
     Collision & Centrality &   $<\beta> $ &    $T$ (MeV) & n & $\chi^2$/ dof  \\\hline
    {\auau} 7.7 GeV& $0-10\%$ & $0.438\pm0.004$& $115\pm2$&0.786 &2.10\\
    & $10-20\%$ &$0.409\pm0.005$   & $118\pm2$& 0.906 & 0.76  \\
    & $20-40\%$ &$0.374\pm0.005$   & $119\pm2$&1.07& 0.78\\
    & $40-80\%$ &$---$   & $---$& $---$   &$---$  \\
    \hline
    {\auau} 11.5 GeV   & $0-10\%$ & $0.435\pm0.004$   & $118\pm2$ & 0.899 & 0.37\\
            & $10-20\%$ &$0.418\pm0.004$   & $120\pm2$         & 0.927&0.35  \\
            & $20-40\%$ &$0.364\pm0.004$   & $129\pm2$        & 1.161&0.58\\
            & $40-80\%$ &$0.240\pm0.04$   & $140\pm2$        & 2.270 & 1.30\\
  \hline
 {\auau} 14.5 GeV   & $0-10\%$ & $0.431\pm0.004$   & $122\pm2$    & 0.891& 0.39\\
            & $10-20\%$ &$0.407\pm0.005$   & $127\pm2$& 0.982& 0.28\\
            & $20-40\%$ &$0.377\pm0.005$   & $132\pm2$ & 1.010&0.40\\
            & $40-80\%$ &$0.234\pm0.005$   & $143\pm2$         &2.488&1.46  \\

  \hline
   {\auau} 19.6 GeV   & $0--10\%$ & $0.451\pm0.005$   & $119\pm2$ & 0.819&0.58\\
            & $10-20\%$ &$0.415\pm0.005$   & $125\pm2$   &0.979&0.35  \\
            & $20-40\%$ &$0.374\pm0.005$   & $129\pm2$& 1.196&0.39\\
            & $40-80\%$ &$0.268\pm0.006$   & $141\pm3$& 1.905&0.35 \\

  \hline
  {\auau} 27 GeV   & $0-10\%$ & $0.465\pm0.003$   & $116\pm3$        & 0.857&0.54\\
            & $10-20\%$ &$0.438\pm0.003$   & $122\pm3$         & 0.933&0.42  \\
            & $20-40\%$ &$0.393\pm0.004$   & $130\pm3$      & 1.120&0.27\\
            & $40-80\%$ &$0.262\pm0.007$   & $149\pm3$         & 2.018&0.40 \\
            
  \hline
   {\auau} 39 GeV   & $0-10\%$ & $0.483\pm0.004$   & $119\pm3$ & 0.695&0.81\\
            & $10-20\%$ &$0.451\pm0.004$   & $125\pm3$          & 0.850&0.35  \\
            & $20-40\%$ &$0.397\pm0.005$   & $136\pm3$         & 1.032&0.37\\
            & $40-80\%$ &$0.289\pm0.006$   & $153\pm3$           & 0.449&0.55\\
            \hline
     {\auau} 62.4 GeV   & $0-10\%$ & $0.497\pm0.007$   & $127\pm4$ &0.671&0.86\\
            & $10-20\%$ &$0.476\pm0.007$   & $133\pm4$ & 0.670&0.85 \\
            & $20-40\%$ &$0.433\pm0.010$   & $144\pm4$ & 0.711&0.80\\
            & $40-80\%$ &$0.340\pm0.018$   & $157\pm4$ & 0.627&0.78\\
            \hline
     {\auau} 200 GeV   & $0-10\%$ & $0.522\pm0.003$   & $116\pm3$ & 0.840&0.78\\
            & $10-20\%$ &$0.504\pm0.004$   & $122\pm3$ &0.867 & 0.61 \\
            & $20-40\%$ &$0.455\pm0.004$   & $129\pm4$ & 1.078&0.54\\
            & $40-80\%$ &$0.345\pm0.005$   & $147\pm4$ & 1.619& 1.33\\
            \hline
     {\pbpb} 2760 GeV   & $0-20\%$ & $0.637\pm0.004$   & $115\pm4$       & 0.698&0.75\\
            & $20-80\%$ &$0.588\pm0.004$   & $125\pm5$ & 0.863&1.15  \\
           
    \hline
 \hline
\end{tabular}
\label{tab2a}
\end{center}} 
\end{table*}
\end{document}